\documentclass[aps,prd,showpacs,twocolumn,superscriptaddress,floatfix]{revtex4}
\usepackage{graphicx,amsmath,amsfonts}

\RequirePackage{xspace}

\newcommand{\BaBarYear}     {08}
\newcommand{\BaBarNumber}  {020}
\newcommand{\SLACPubNumber} {13301}

\newcommand{\BaBarType}      {PUB}  

\input babarsym

\def\BR#1#2{\ensuremath{{\cal B}(#1\to#2)\xspace}}
\def\GG#1#2{\ensuremath{\Gamma(#1\to#2)\xspace}}

\begin{document}

\noindent\babar-\BaBarType-\BaBarYear/\BaBarNumber \\  
SLAC-PUB-\SLACPubNumber\\
arXiv:0807.2014 [hep-hex]\\

\smallskip

\title{
\large \bf \boldmath
Study of hadronic transitions between $\Upsilon$ states and
observation of $\Upsilon(4S)\to\eta\Upsilon(1S)$ decay
}

\begin{abstract}

We present a study of hadronic transitions between $\Upsilon(mS)$ ($m=4,3,2)$ and  $\Upsilon(nS)$ ($n=2,1$) resonances 
based on 347.5\invfb of data taken with the \babar\ detector at the
\pep2\ storage rings. We report the first observation 
of  $\FourS\to\eta\OneS$ decay with a branching fraction
${\cal B}(\FourS\to\eta\OneS)=(1.96\pm0.06_{stat}\pm0.09_{syst})\times10^{-4}$
and measure the ratio of partial widths
 $\GG{\FourS}{\eta\OneS}/\GG{\FourS}{\pipi\OneS}=2.41\pm 0.40_{stat}\pm 0.12_{syst}$.
We set 90\% CL upper limits on the ratios
$\GG{\TwoS}{\eta\OneS}/\GG{\TwoS}{\pipi\OneS}<5.2\times10^{-3}$ and 
$\GG{\ThreeS}{\eta\OneS}/\GG{\ThreeS}{\pipi\OneS}<1.9\times10^{-2}$.
We also present new measurements of the ratios
$\GG{\FourS}{\pipi\TwoS}/\GG{\FourS}{\pipi\OneS}=1.16\pm 0.16_{stat}\pm 0.14_{syst}$ and
$\GG{\ThreeS}{\pipi\TwoS}/\GG{\ThreeS}{\pipi\OneS}=0.577\pm 0.026_{stat}\pm 0.060_{syst}$.

\end{abstract}

%
\author{B.~Aubert}
\author{M.~Bona}
\author{Y.~Karyotakis}
\author{J.~P.~Lees}
\author{V.~Poireau}
\author{E.~Prencipe}
\author{X.~Prudent}
\author{V.~Tisserand}
\affiliation{Laboratoire de Physique des Particules, IN2P3/CNRS et Universit\'e de Savoie, F-74941 Annecy-Le-Vieux, France }
\author{J.~Garra~Tico}
\author{E.~Grauges}
\affiliation{Universitat de Barcelona, Facultat de Fisica, Departament ECM, E-08028 Barcelona, Spain }
% ...
\collaboration{The \babar\ Collaboration}
\noaffiliation


\maketitle

\setcounter{footnote}{0}

\section{\boldmath Introduction}
\label{sec:intro}
Hadronic transitions between bound states of heavy quarkonia~\cite{Brambilla:2004wf} 
are generally studied using the QCD multipole expansion model (QCDME)~\cite{Kuang-Rev}.
This succeeds in explaining the relative rates of the $\psi(2S)\to\eta \jpsi$ and 
$\psi(2S)\to\pi\pi \jpsi$ transitions and
the $\pi\pi$ invariant mass distributions in $\psi(2S)\to\pi\pi \jpsi$,
$\TwoS\to\pi\pi\OneS$, $\ThreeS\to\pi\pi\TwoS$ and the recently observed
$\FourS\to\pipi\OneS$ decays~\cite{Aubert:2006bm,Sokolov:2006sd}. 
Until recently the only feature that QCDME could not explain 
was the dipion invariant mass distribution in the $\ThreeS\to\pi\pi\OneS$ transition~\cite{Butler:1993rq}, 
for which a number of possible explanations have been proposed~\cite{Boh}.  
The  dipion invariant mass distribution in $\FourS\to\pipi\TwoS$~\cite{Aubert:2006bm} is also in disagreement with the QCDME prediction and was not predicted
either by the alternative explanations proposed for the 
$\ThreeS\to\pipi\OneS$. This 
implies that additional experimental 
input is needed to understand hadronic transitions. 
In QCDME the gluon radiation from a heavy $q\bar q$ bound state is
calculated in terms of chromo-electric and chromo-magnetic fields,
in analogy to electromagnetism.  
Transitions between colorless hadrons require the emission of at least two gluons.
The $\Upsilon(mS)\to\pi\pi\Upsilon(nS)$ transitions ($m^3S_1\to \pi\pi~n^3S_1$ in spectroscopic notation~\cite{my-FN}) are E1E1, i.e.
transitions where both gluons are in an E1 state.
The decays $\Upsilon(mS)\to\eta\Upsilon(nS)$  ($m^3S_1\to\eta~ n^3S_1$) 
proceed either via E1M2 or M1M1 transitions; the
E1M2 transition is expected to dominate.
The $\bbbar$ system offers unique opportunities: 
there are five known $m^3S_1\to \pi\pi~n^3S_1$ transitions and also four
kinematically allowed transitions involving an $\eta$ meson.  
Of the latter only the $\TwoS\to\eta\OneS$ has been recently observed by
CLEO~\cite{He:2008xk}, with a branching fraction ${\cal B}(\TwoS\to\eta\OneS)=
(2.1^{+0.7}_{-0.6}\pm0.5)\times10^{-4}$.

In this paper we present improved measurements of the 
$\FourS\to\Upsilon(nS)$ transitions, a search for $\Upsilon(mS)\to\eta\Upsilon(1S)$ and
new measurements of $\ThreeS\to\pipi\Upsilon(nS)$ and $\TwoS\to\pipi\OneS$
partial widths. We also measure the ratios of partial widths
$\GG{\Upsilon(mS)}{\eta\Upsilon(1S)}/\GG{
\Upsilon(mS)}{\pipi\Upsilon(1S)}$ and $\GG{\Upsilon(mS)}{\pipi\Upsilon(2S)}/\GG{
\Upsilon(mS)}{\pipi\Upsilon(1S)}$ $(m=3,4)$, for which a number of systematic uncertainties cancel.

The $\Upsilon(mS)\to\pipi\Upsilon(nS)$ and $\Upsilon(mS)\to\eta\Upsilon(nS)$ transitions, 
denoted by $mS\to\pi\pi~nS$ and $mS\to\eta~nS$, respectively, are studied
by reconstructing the $\Upsilon(nS)$ mesons via their leptonic decay to $\mumu$ or $\epem$.
The $\eta$ meson is reconstructed via its $\pipi\pi^0$ decay.
With the choice of this particular $\eta$ decay mode all final states contain 
the same charged particles, resulting in larger cancellations of the systematic 
uncertainties for the ratios of partial widths.
Events where the $\eta$ decays to $\gamma\gamma$ are not considered in this work
because the $\ell\ell\gamma\gamma$ final state has a smaller signal to background ratio 
than the $\ellell\pipi\pi^0$ final state.

\section{\boldmath Data samples and detector}
\label{sec:outline}

We search for \FourS\ hadronic transitions using a sample of
$(383.2\pm4.2)\times10^6$  \FourS\ 
decays corresponding to an integrated luminosity, $L^{int}_{on}$, of 347.5$\,\invfb$ acquired  
near the peak of the \FourS resonance (``on-peak'',  nominal center-of-mass energy, $\sqrt{s}$ of about 10.58~\gev) with the \babar\ detector at the \pep2\
asymmetric-energy $\epem$ storage rings at SLAC. 
In addition, a data sample corresponding to $L^{int}_{off}=36.6\,\invfb$, collected approximately 40$\,\mev$ below 
the resonance  (``off-peak'') is used to study some of the backgrounds. 
Decays of \ThreeS\ and \TwoS\  are studied in events recorded ``on-peak'' and selected with an initial state radiation (ISR) photon.
The ISR photon, preferentially emitted at small angle along the beam direction,
is not required to be detected.
 
The \babar\ detector is described in detail elsewhere~\cite{babar-detector}.
Charged-particle momenta are measured in a tracking system consisting 
of a five-layer double-sided silicon vertex tracker (SVT) and a 
40-layer central drift chamber (DCH), both embedded in a 1.5-T axial 
magnetic field. Charged-particle identification is based on
the specific energy loss measured in the SVT and DCH,
and on a measurement of
the photons produced in the fused-silica bars of the
ring-imaging Cherenkov detector (DIRC).
A CsI(Tl) electromagnetic 
calorimeter (EMC) is used to detect and identify photons and 
electrons, while muons are identified in the instrumented 
flux return of the magnet (IFR).

\begin{figure*}[!htb]
\begin{center}
\begin{tabular}{cc}
\includegraphics[scale=0.43]{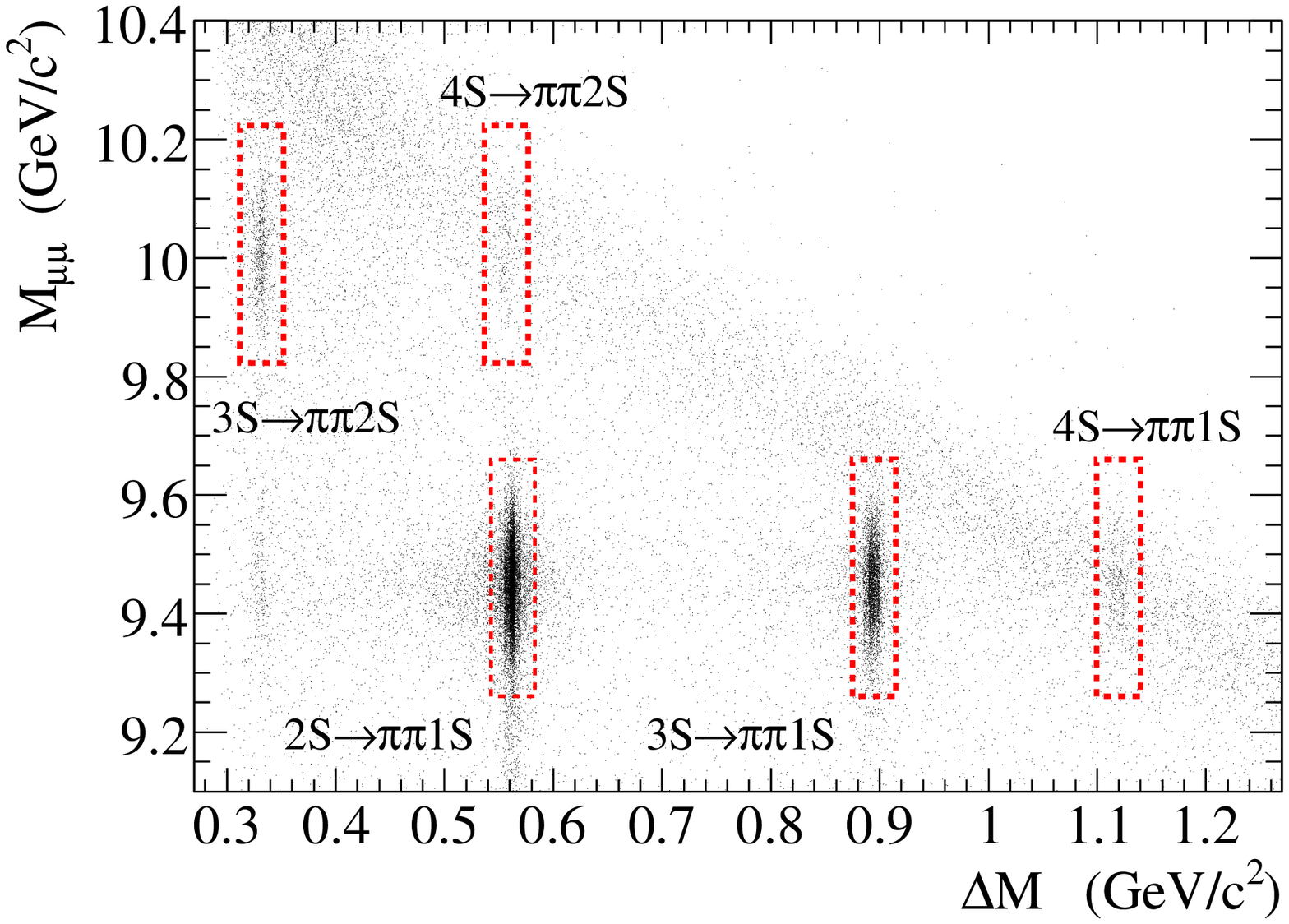}&
\includegraphics[scale=0.43]{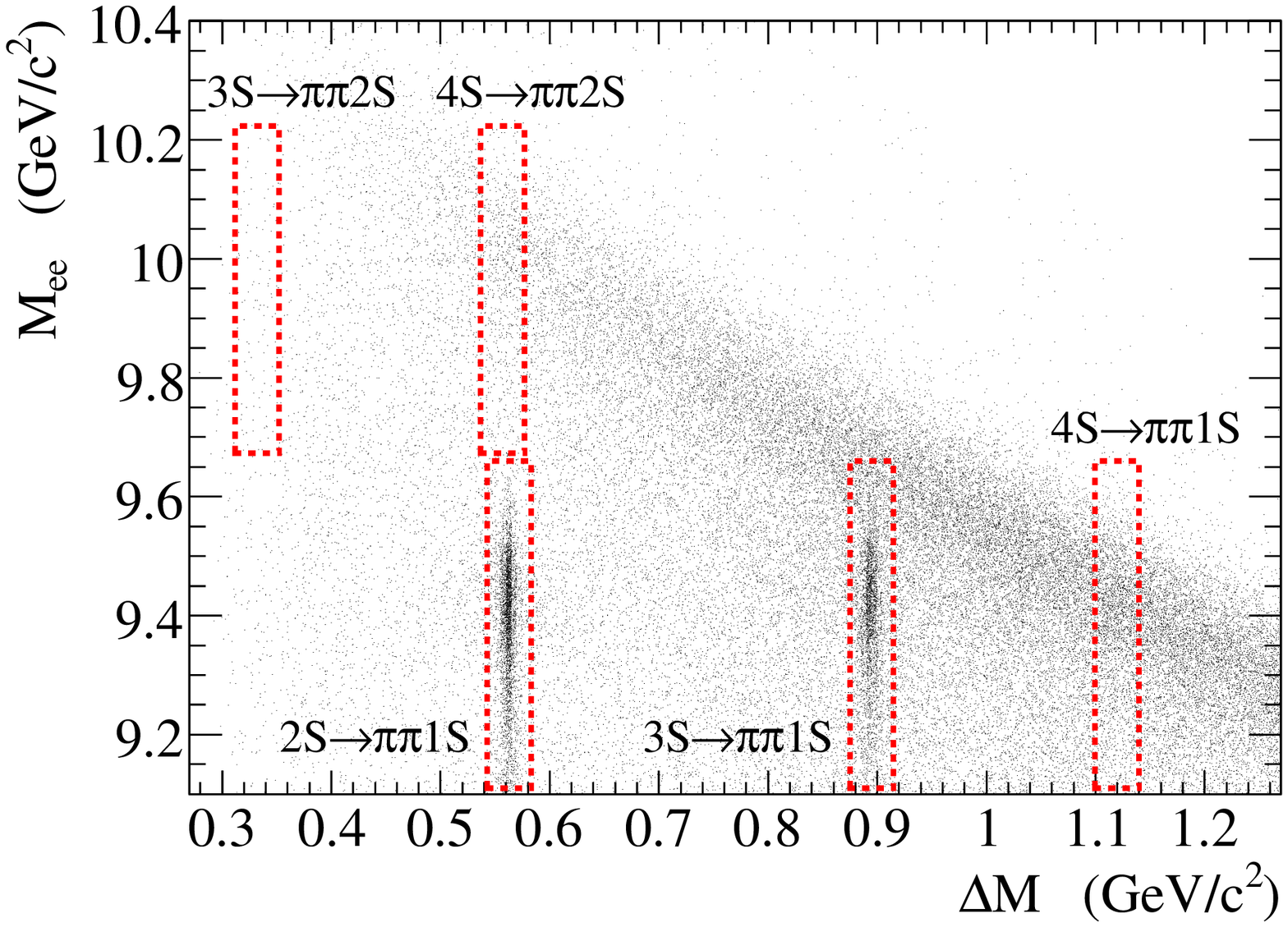}\\
\end{tabular}
\end{center}
\caption{\label{fig:DeltaMvsMll}$M_{\ell\ell}$  vs $\Delta M$ distributions of candidates after
the preliminary selection for the $\mu\mu$ (left) and $ee$ (right) samples.
Dashed lines delimit the signal boxes for $mS\to\pi\pi~nS$ transitions.
The cluster of events in the lower left corner is due to $\ThreeS\to\pipi\TwoS$ where
the $\TwoS$ subsequently decays to $\OneS\,X$.}
\end{figure*}

Simulated Monte Carlo (MC) events are generated 
using the EvtGen package~\cite{EVTGEN}.
The angular distribution of generated dilepton decays incorporates 
the $\Upsilon(nS)$ polarization, 
while dipion transitions are generated according to phase space.
In the simulation of $mS\to \eta~1S$ we use the
angular distribution dictated by the quantum numbers for a vector decay to a pseudoscalar and a vector.
Secondary photon emission is taken into account in the simulation of  $\Upsilon(mS)$ produced
in ISR. Simulated events are passed through a detector simulation based on GEANT4~\cite{GEANT},
and analyzed in the same manner as data.

\section{\boldmath Event selection}
\label{sec:sel}

The events of interest have a lepton pair from the decay of the $\Upsilon(nS)$ resonance
of invariant mass, $M_{\ell\ell}$, compatible with the known mass values of the
$\Upsilon(nS)$~\cite{PDG07}, $M(nS)$, and a pair of oppositely charged pions.

The signature for $mS\to\pi\pi\; nS$ transition events is an invariant mass difference
$\Delta M=M_{\pi\pi\ell\ell}-M_{\ell\ell}$ compatible with the difference of the masses of the two $\Upsilon$ resonances, $M(mS)-M(nS)$, where  $M_{\pi\pi\ell\ell}$ is the $\pipi\ellell$ invariant mass.

The $mS\to\eta~nS$ events have two additional photons from
the $\pi^0$ decay, a $\pipi\pi^0$ invariant mass, $m_{3\pi}$,
compatible with the known $\eta$ mass, $M(\eta)$, and an invariant mass
difference, $\Delta M_\eta=M_{3\pi\ell\ell}-M_{\ell\ell}-m_{3\pi}$ compatible with $M(mS)-M(nS)-M(\eta)$, where $M_{3\pi\ell\ell}$ is the 
$\pipi\pi^0\ellell$ invariant mass.

The  r.m.s. widths of the reconstructed $M_{\ell\ell}$, $m_{3\pi}$, $\Delta M$,
and $\Delta M_{\eta}$ distributions are of the order of 
$75\,\mevcc$, $12\,\mevcc$, $ 7\,\mevcc$ and 
$ 10\,\mevcc$, respectively.

Events in the data sample with $M_{\ell\ell}$ within
350~\mevcc of the known $M(nS)$ values and 
$\Delta M$ within 60~\mevcc of 
the values expected for any of the $mS\to \pi\pi\;nS$ transitions were 
not examined until the 
event selection criteria were finalized.  Events outside these regions
were used to understand the background. Simulated MC events were used
to model the signal.

Candidate events have at least 4 charged tracks with a polar angle $\theta$ within
the fiducial volume of the tracking system (0.41$<\theta<$2.54~rad).
Each lepton candidate is required to have a center-of-mass momentum between 
4.20\gevc and 5.25\gevc. At least one of the muons of $\Upsilon(nS)\to\mumu$ candidates must 
be compatible with the muon hypothesis based on the energy deposited in the EMC 
and the hit pattern in the IFR along the track trajectory.
Similarly at least one of the electrons of $\Upsilon(nS)\to\epem$ candidates 
must be compatible with the electron 
hypothesis based on the energy deposit in the EMC, the ratio of energy in the 
EMC to the track momentum,
and the energy loss in the detector material. We require  
$M_{\mu\mu}$[$M_{ee}$] to be within $\pm$200 [$-$350,+200]~\mevcc of
the nominal \OneS\ or \TwoS\ mass. The asymmetric cut in the \epem\ sample is due to 
bremsstrahlung, which causes a long tail  in the reconstructed
$M_{ee}$ distribution at low invariant masses and that is partially recovered 
by an algorithm that combines the energy of electron tracks with
the energy of nearby photons.

Pairs of oppositely charged tracks, not identified
as electrons and whose Cherenkov angle in the DIRC, when measured, is within $3\,\sigma$ of the value 
expected for a pion, are selected to form a dipion candidate.
The dilepton and the dipion are constrained to a common vertex and the vertex fit is required to 
have a $\chi^2$ probability larger than $10^{-3}$.

\begin{table}[htb]
\caption{\label{tab:effs} Selection efficiencies for all studied transitions, separately 
for $\Upsilon(nS)\to\mumu$ and $\epem$ as determined by MC simulation. For the
$mS\to\pi\pi\;nS$ transitions we quote both the efficiency averaged over
phase space, $\varepsilon_{PS}$, and the effective efficiency, $\varepsilon_{eff}$, calculated according to Eq.~\ref{effdef}.}
\begin{center}
\begin{tabular}{l|cccc}
\hline
\hline
 Transition & \multicolumn{4}{c}{Selection efficiency (\%)} \\
 &  \multicolumn{2}{c}{$\mu\mu$} &  \multicolumn{2}{c}{$ee$} \\
 &  $\varepsilon_{PS}$ & $\varepsilon_{eff}$  &  $\varepsilon_{PS}$ &
 $\varepsilon_{eff}$  \\
\hline
$2S\to\pi\pi\;1S$  &34.46$\pm$0.05 & 36.62$\pm$0.08 & 11.17$\pm$0.03 & 11.45$\pm$0.14 \\
$3S\to\pi\pi\;1S$  &41.23$\pm$0.05 & 34.18$\pm$0.20 & 24.48$\pm$0.05 & 23.96$\pm$0.24 \\
$3S\to\pi\pi\;2S$  &14.76$\pm$0.04 & 17.2 $\pm$0.6  & $\approx 0$    & -- \\
$4S\to\pi\pi\;1S$  &41.53$\pm$0.23 & 44.2 $\pm$1.2  & 18.04$\pm$0.18 & 19.7 $\pm$2.4 \\ 
$4S\to\pi\pi\;2S$  &32.69$\pm$0.22 & 30.2 $\pm$0.8  &  6.17$\pm$0.12 & 7.9 $\pm$3.4 \\
\hline
 &  \multicolumn{2}{c}{$\varepsilon$} &  \multicolumn{2}{c}{$\varepsilon$} \\
\hline
$2S\to\eta~1S$   &  \multicolumn{2}{c}{8.25$\pm$0.09} &   \multicolumn{2}{c}{$\approx 0$}   \\
$3S\to\eta~1S$   &  \multicolumn{2}{c}{9.42$\pm$0.10} &   \multicolumn{2}{c}{3.91$\pm$0.06}  \\
$4S\to\eta~1S$   & \multicolumn{2}{c}{10.07$\pm$0.10} &   \multicolumn{2}{c}{3.77$\pm$0.06} \\
\hline
\hline
\end{tabular}
\end{center}\end{table}

A large fraction of the remaining background is due to $\epem\gamma$ and $\mumu\gamma$ events where a 
photon converts
in the detector material and the leptons are reconstructed as pions.
To reduce this background we
reject events where the opening angle of the charged pion candidates in the laboratory reference frame has $\cos{\theta_{\pipi}}>0.95$, 
or where the invariant mass of the charged tracks associated with the pion 
candidates, calculated assuming the $e^\pm$ mass hypothesis, satisfies $m_{conv}<50\,\mevcc$. 
The distribution of $M_{\ell\ell}$ vs $\Delta M$for candidate events after the preliminary
selection is shown in Fig.~\ref{fig:DeltaMvsMll}.

In the case of $\FourS\to \Upsilon(nS)$ transitions the background is larger and 
the expected signal smaller. For this reason  we further restrict our selection to events where 
at least one of the two charged pions has a 
transverse momentum greater than $100\,\mevc$, $m_{conv}>100\,\mevcc$,  
and the polar angle in the laboratory system of the $e^-$ from $\Upsilon(nS)\to\epem$ is larger 
than 0.7~rad, to reject radiative Bhabha events.

Events with at least two  candidate photons of $E_\gamma>50~\mev$ and invariant mass
$110 < m_{\gamma\gamma}< 150\,\mevcc$ are considered to be $\eta~nS$ candidates if the
$\pipi\pi^0$  invariant mass  is within $35\,\mevcc$  of the 
known $\eta$ mass. To suppress possible cross-feed from the high statistics $mS\to \pi\pi\; nS$ 
transitions we require that $mS\to\eta~nS$ candidates have $\Delta M$ more  
than 20~\mevcc ($\approx 3\sigma$) from any of the known $M(mS)-M(nS)$ values.

We select  \TwoS\ and \ThreeS\ states produced via ISR, 
requiring that the momentum of the reconstructed $\ellell\pipi[\pi^0]$ in the
center of mass rest frame, $p^*_{cand}$, is  within $\pm 150\,\mevc$ 
of the expected value of $\left(s-M^2(mS)\right)/(2\sqrt{s})$. For \FourS\ decays
$p^*_{cand}$ is required to be $<200\,\mevc$.

The average efficiency for each of the transitions is given in Table~\ref{tab:effs}.
The efficiency for the $\pipi[\pi^0]\epem$ final state is in all 
cases smaller than for the $\pipi[\pi^0]\mumu$ final state due to a trigger-level inefficiency 
introduced by the prescaling of Bhabha scattering events, whose signature is given by two 
electrons of large invariant mass and no additional charged track of transverse momentum 
greater than 250~\mevc. 
Because of the limited phase-space available in the $2S\to~\eta ~1S$ and 
$3S\to~\pi\pi~2S$ decays, the momentum of the charged pions is always below the threshold, thus the efficiency for these two transitions is nearly zero when  the  
$\Upsilon(nS)$ decays to $\epem$.

\begin{figure*}[!htb]
\begin{center}
\begin{tabular}{ccc}
\includegraphics[scale=0.29]{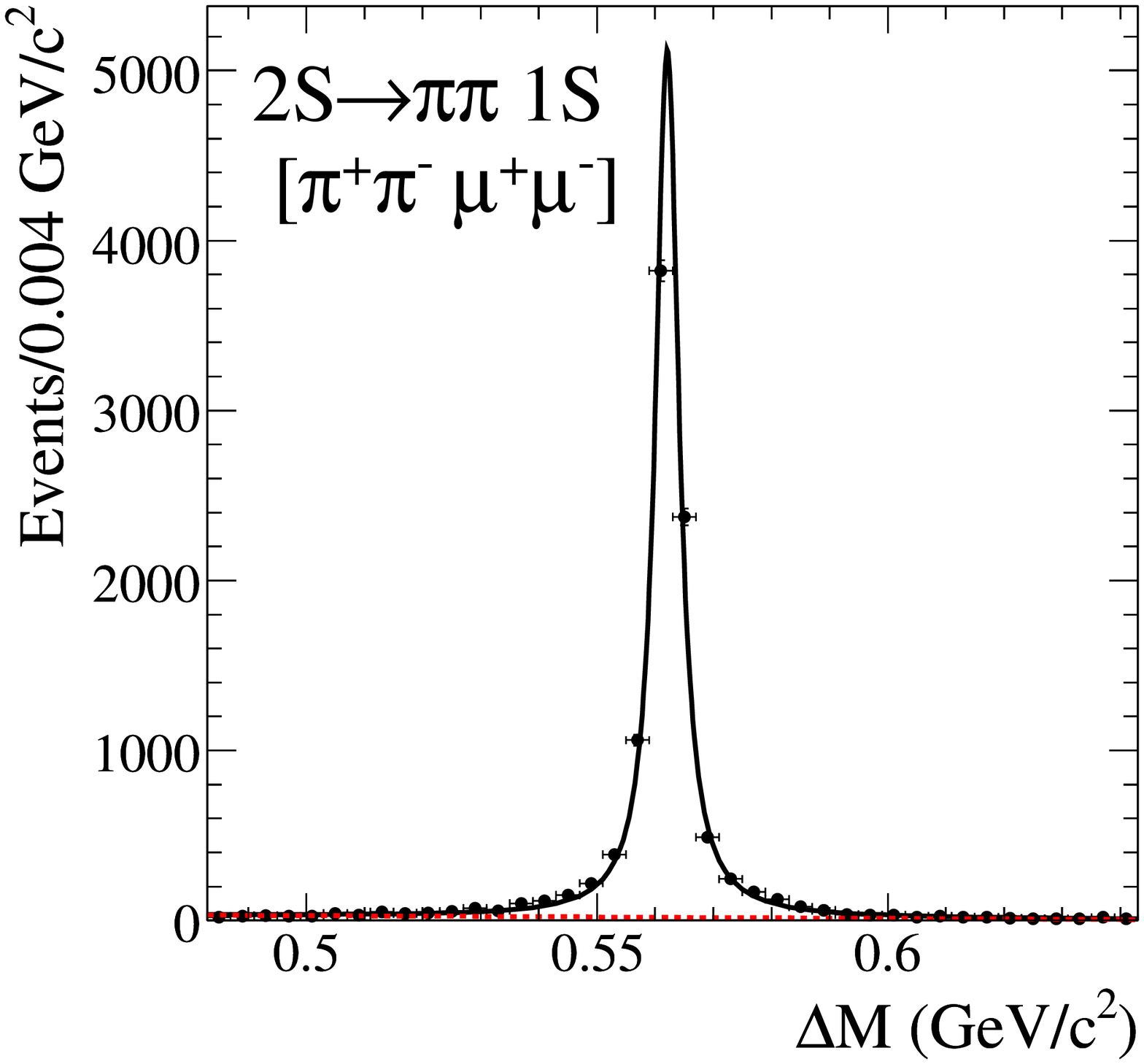}&
\includegraphics[scale=0.29]{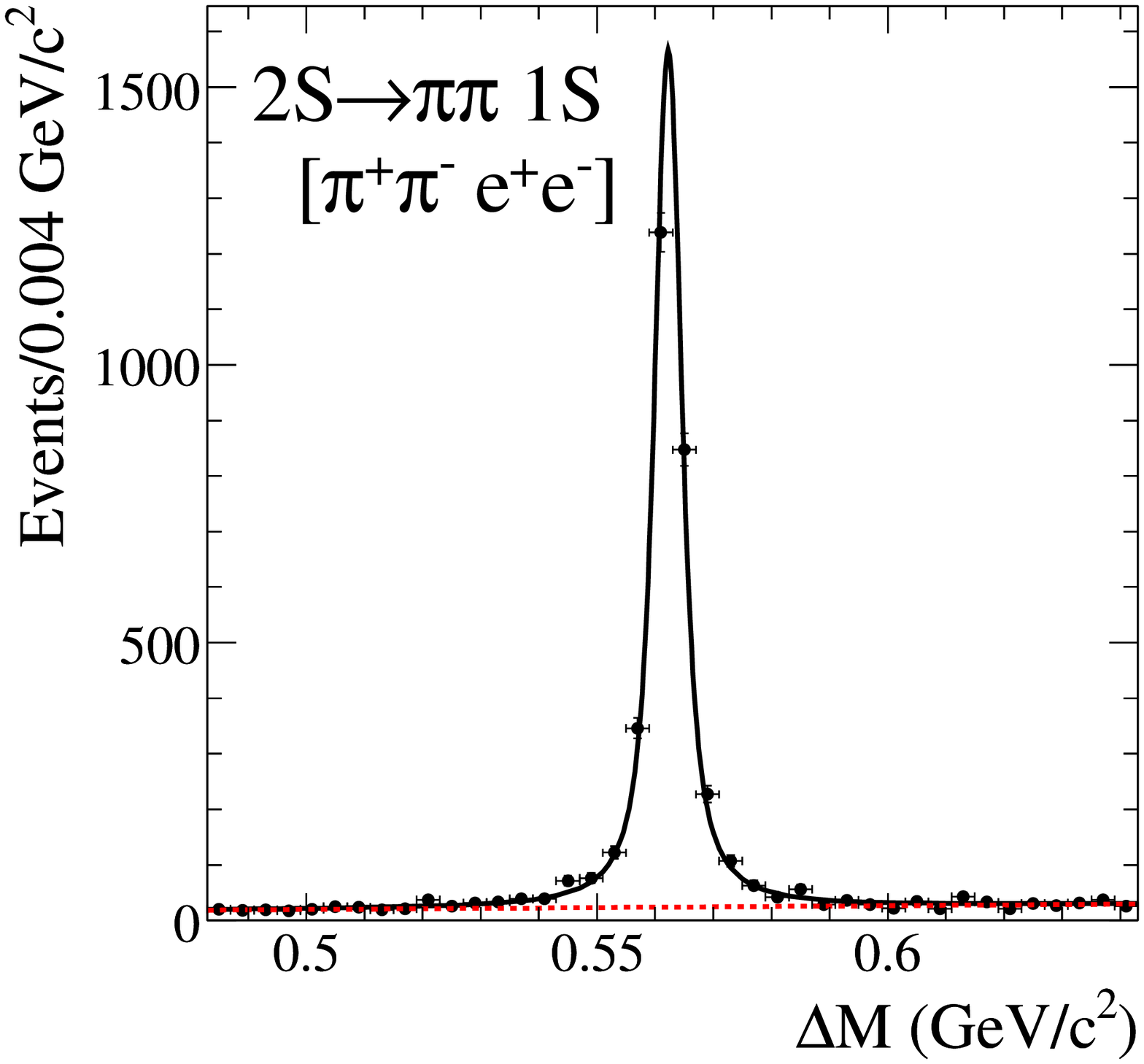}&
\includegraphics[scale=0.29]{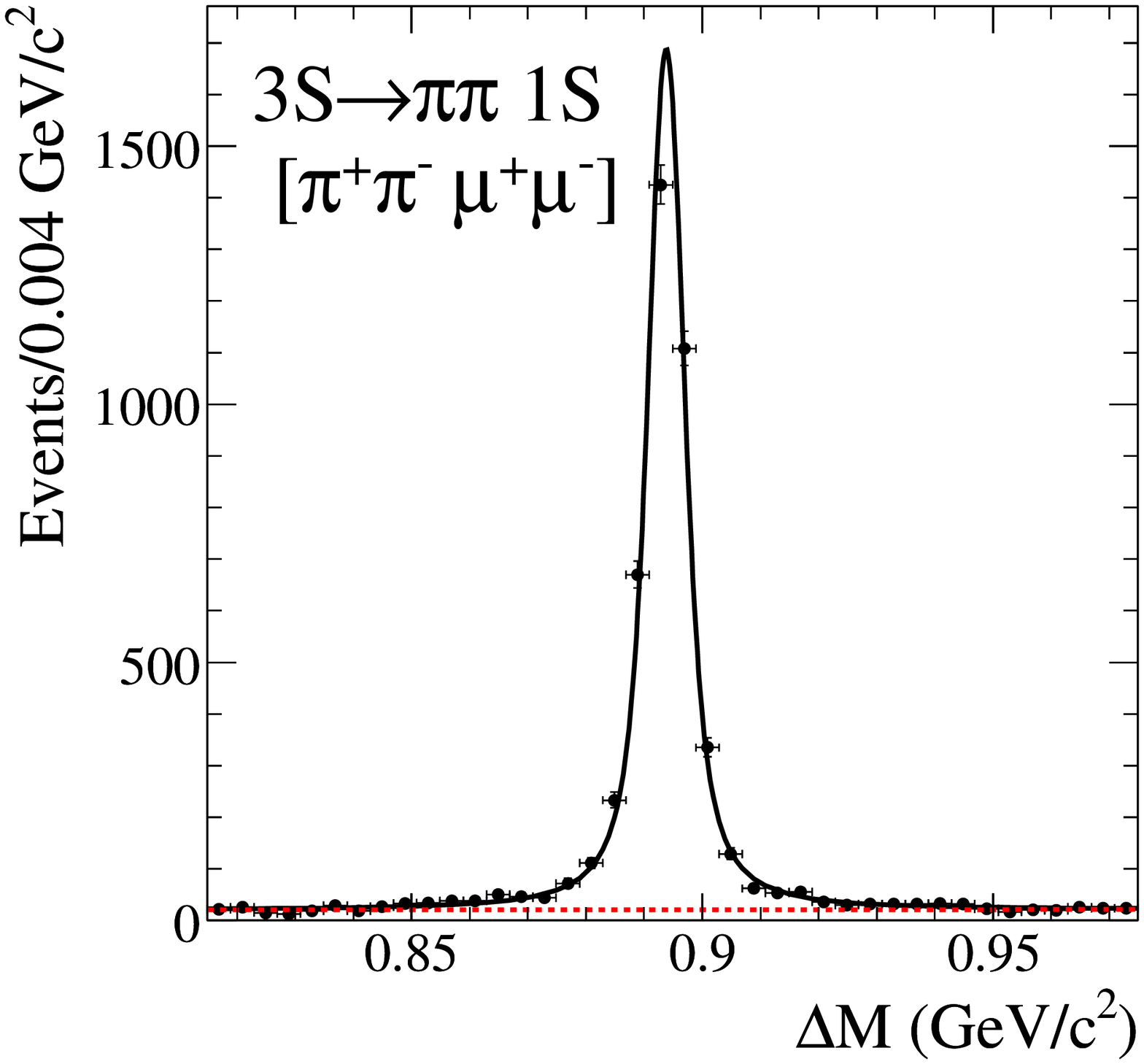}\\
\includegraphics[scale=0.29]{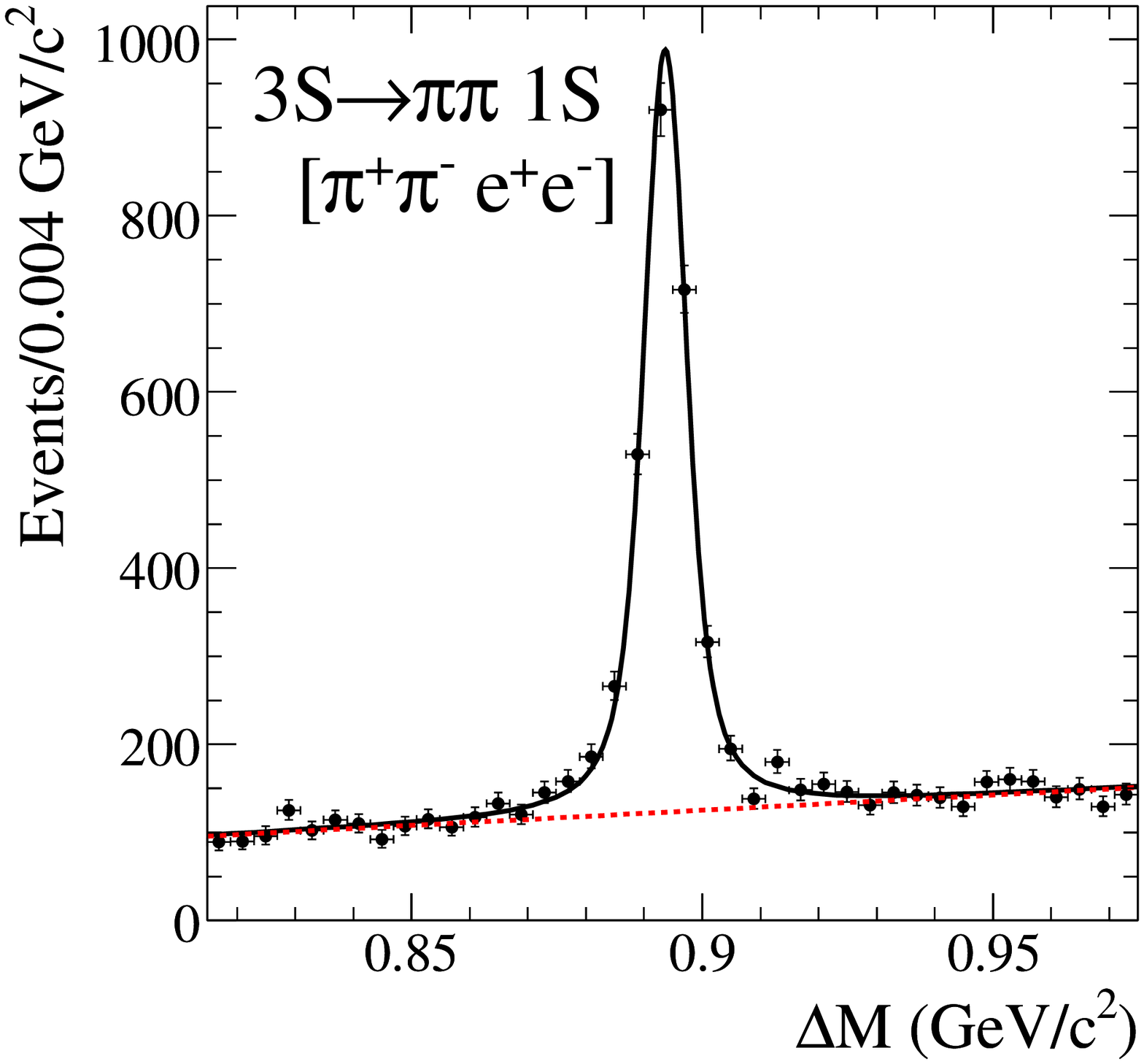}&
\includegraphics[scale=0.29]{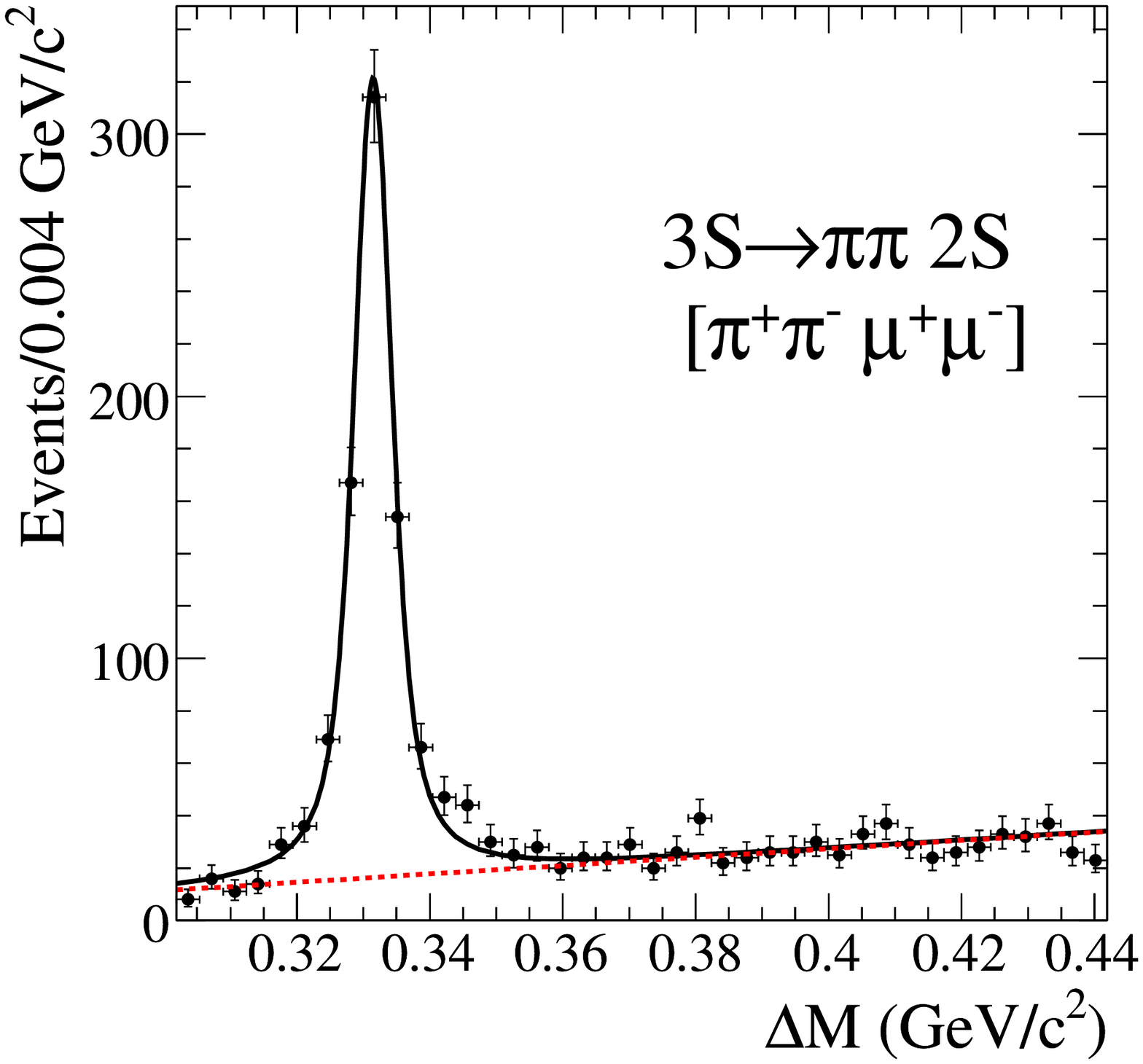}&
\includegraphics[scale=0.29]{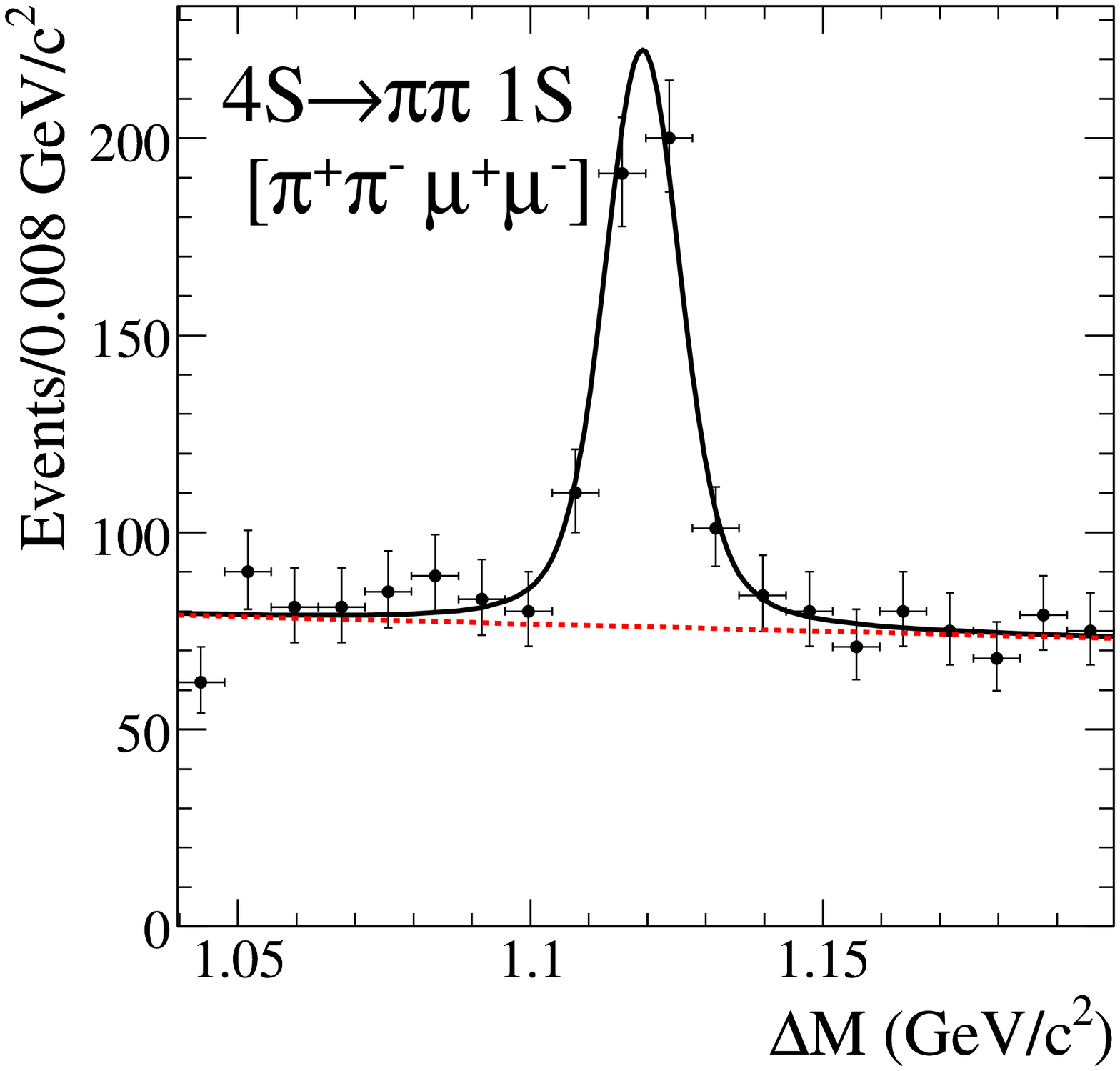}\\
\includegraphics[scale=0.29]{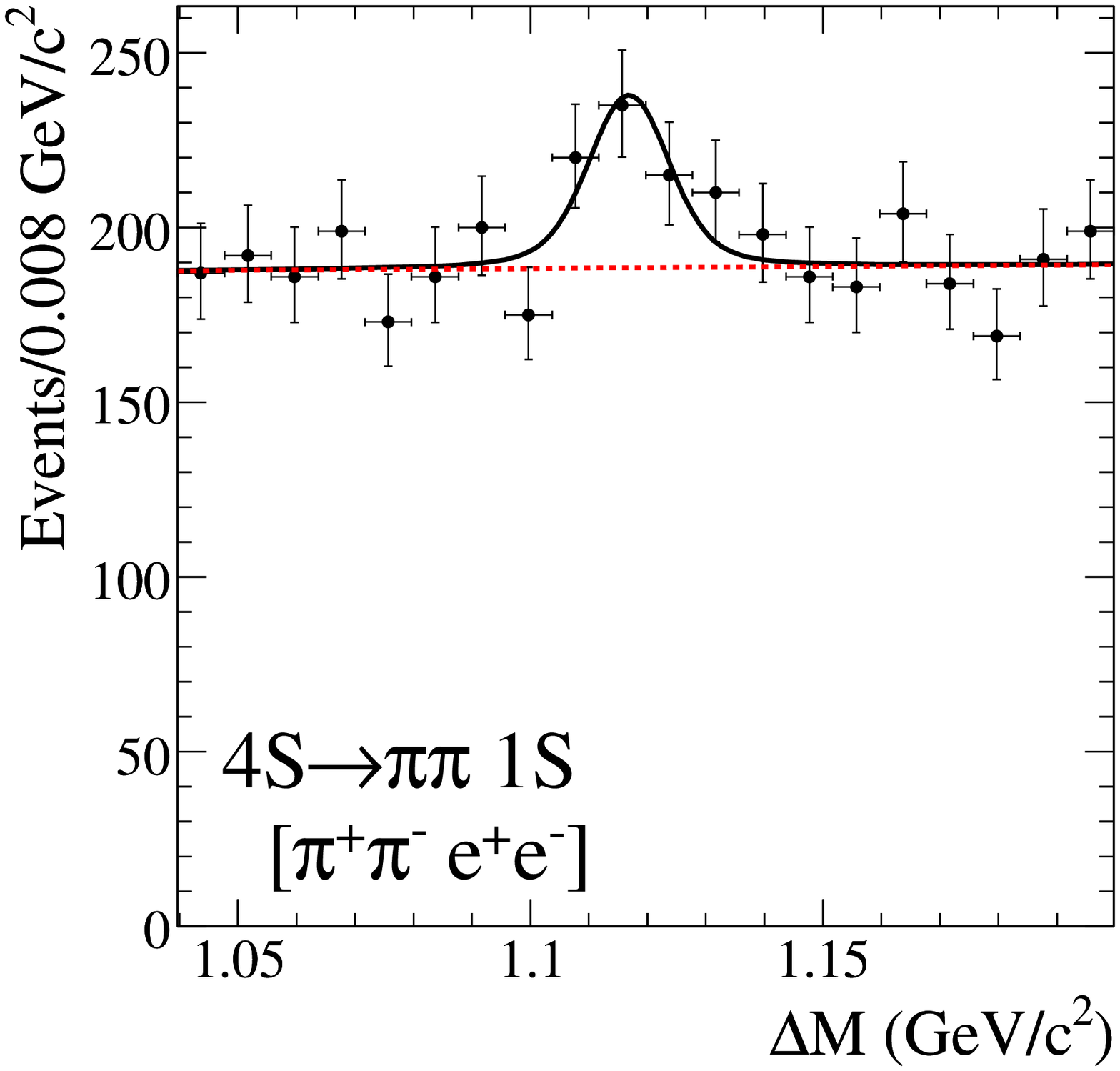}&
\includegraphics[scale=0.29]{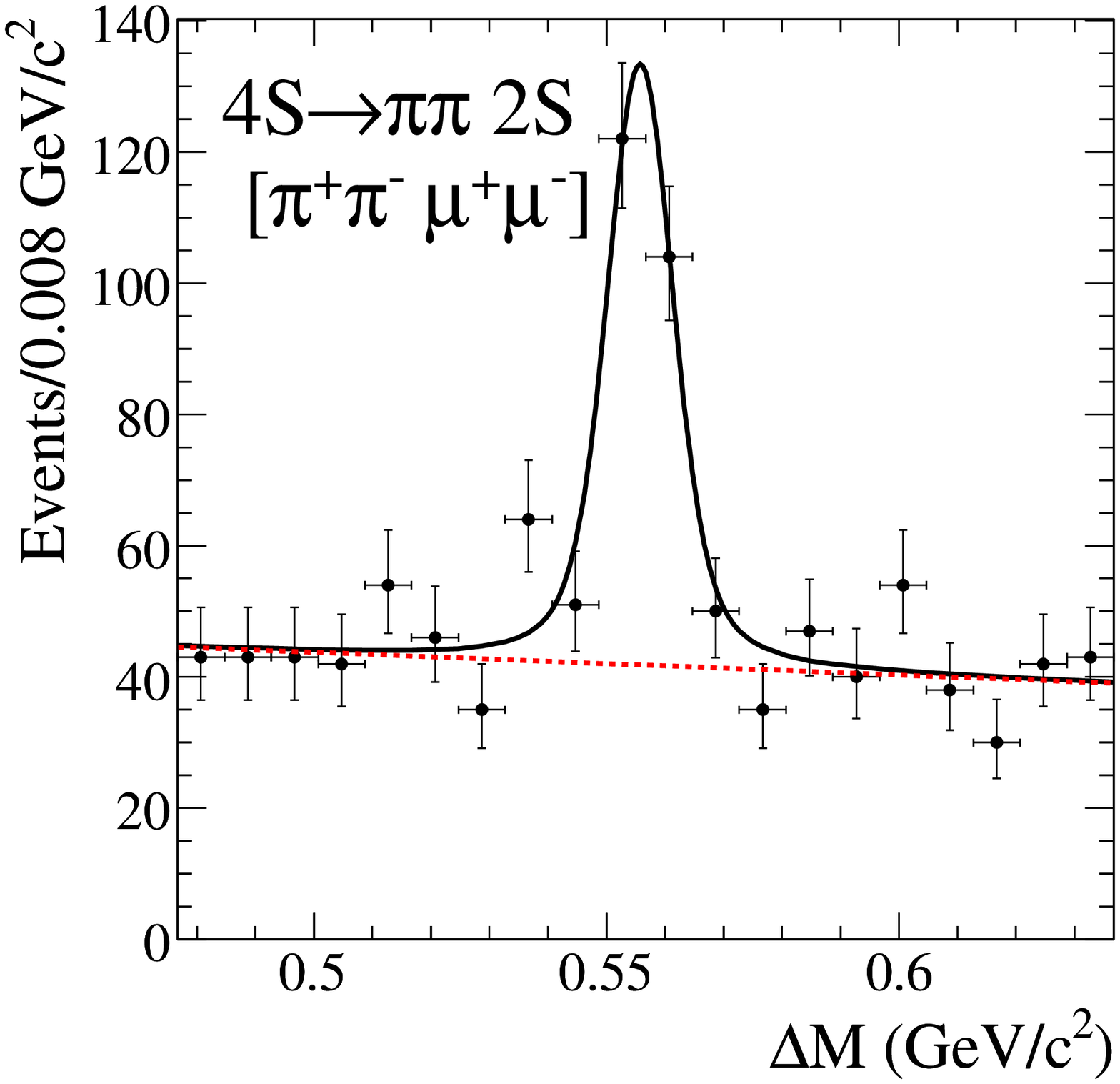}&
\includegraphics[scale=0.29]{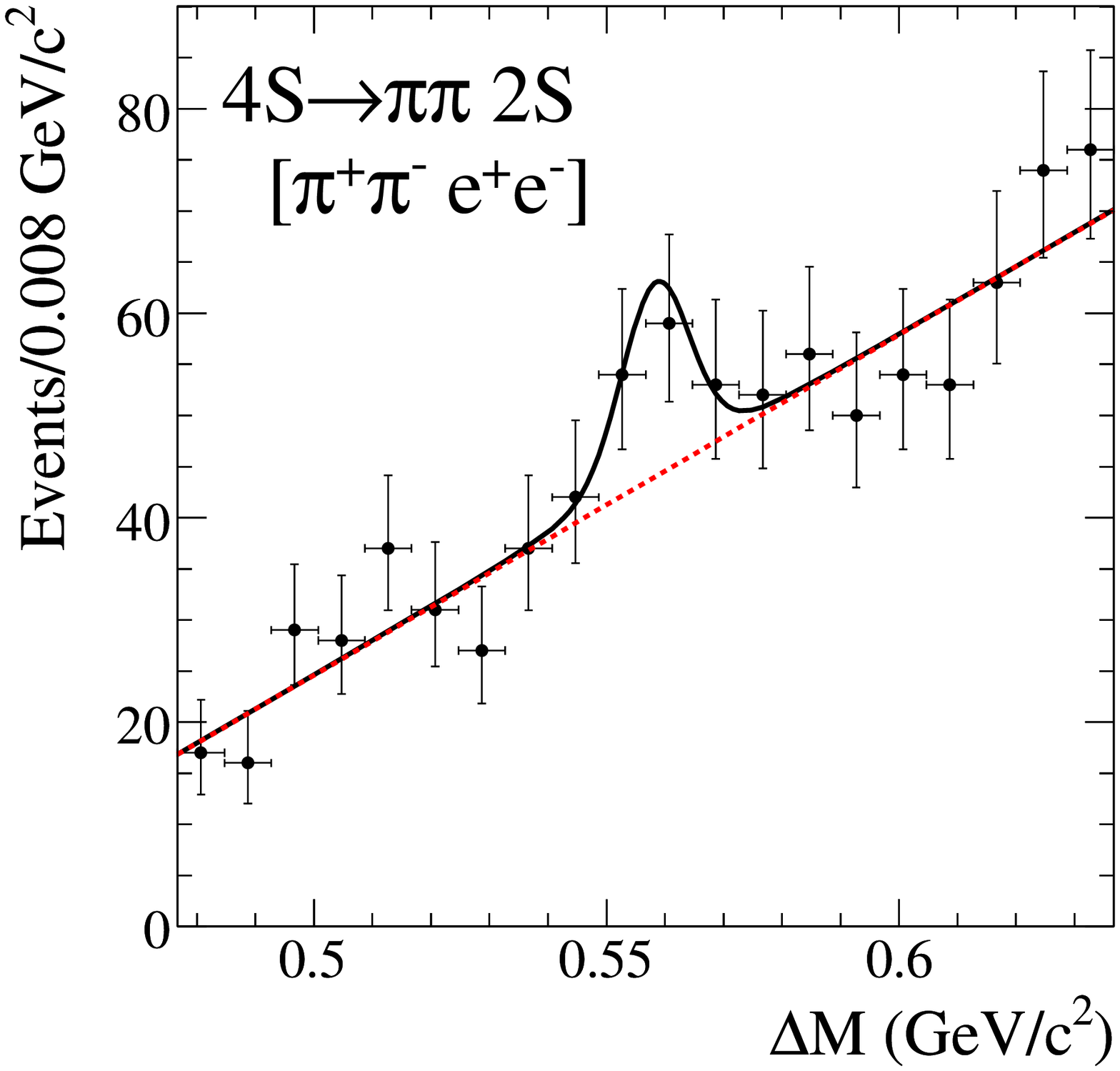} 
\end{tabular}
\end{center}
\caption{\label{fig:UpsPipi}$\Delta M$ distributions of events in the final sample for
the $mS\to\pi\pi\;nS$ transitions.
Data are shown as crosses. The solid lines are the best fit to the data and are
only for illustration purposes: they are performed
using the signal PDF described in the text with resolution parameters fixed to the values determined on MC events.  
Dashed lines show the background contribution.}\end{figure*}

\section{\boldmath Signal yields}
\label{sec:sigYield}
\subsection{\boldmath $\Upsilon(mS)\to\pipi\Upsilon(nS)$}
The $\Delta M$ distributions of events in the final sample for
the $mS\to\pi\pi~nS$ transitions are shown in Fig.~\ref{fig:UpsPipi}.

We determine the efficiency corrected signal yield for the $mS\to\pi\pi\,nS$ transitions without 
any assumption on the angular distributions of the decays.
We divide the $2S\to\pi\pi~1S$ and $3S\to\pi\pi~ 1S$ samples into
10$\times$6 bins of $m_{\pi\pi}$ and
$\cos{\theta_{h}}$, where  $m_{\pi\pi}$ is the $\pipi$ invariant mass and  $\theta_{h}$ is the helicity angle of the $\pi^+$, defined 
as the angle between the $\pi^+$ direction in the $\pi\pi$ rest frame and the $\pi\pi$ direction in the 
candidate $\Upsilon(mS)$ rest frame.
The $3S\to\pi\pi\;2S$ and $4S\to\pi\pi\; nS$ samples are divided into 6$\times$4 bins of  $m_{\pi\pi}$ and
$\cos{\theta_{h}}$.

The signal yield in each bin is determined by a fit to the 
$\Delta M$ distribution,
by maximizing the unbinned extended likelihood to the sum of a background
probability density function (PDF) and a signal PDF. The signal PDF is parametrized by a Voigtian function (convolution of a Lorentzian with a Gaussian function), that is found to describe well the measured $\Delta M$ distribution for simulated events.
The background is parametrized by a linear function.
The resolution parameters for the signal PDF are fixed to the values determined by the simulation, thus
the free parameters in the fits for bin $i$ are: 
$\Delta M_{sig}^{i}$, the peak position of the signal distribution, $N_{sig}^{i}$ and $N_{bkg}^i$, the number of signal and background events, 
and the background shape parameters.
The efficiency corrected signal yield for each $mS\to\pi\pi\,nS$ transition 
is then obtained as 
\begin{equation}
N_{corr}=\sum_{i=1}^{nbins}N_{sig}^i/\varepsilon_i,
\end{equation}
where $nbins$ is the number of bins 
(60 or 24) and $\varepsilon_i$ is the efficiency in each bin
determined from MC simulation.

\begin{figure}[!htb]
\begin{center}
\begin{tabular}{cc}
\includegraphics[scale=0.22]{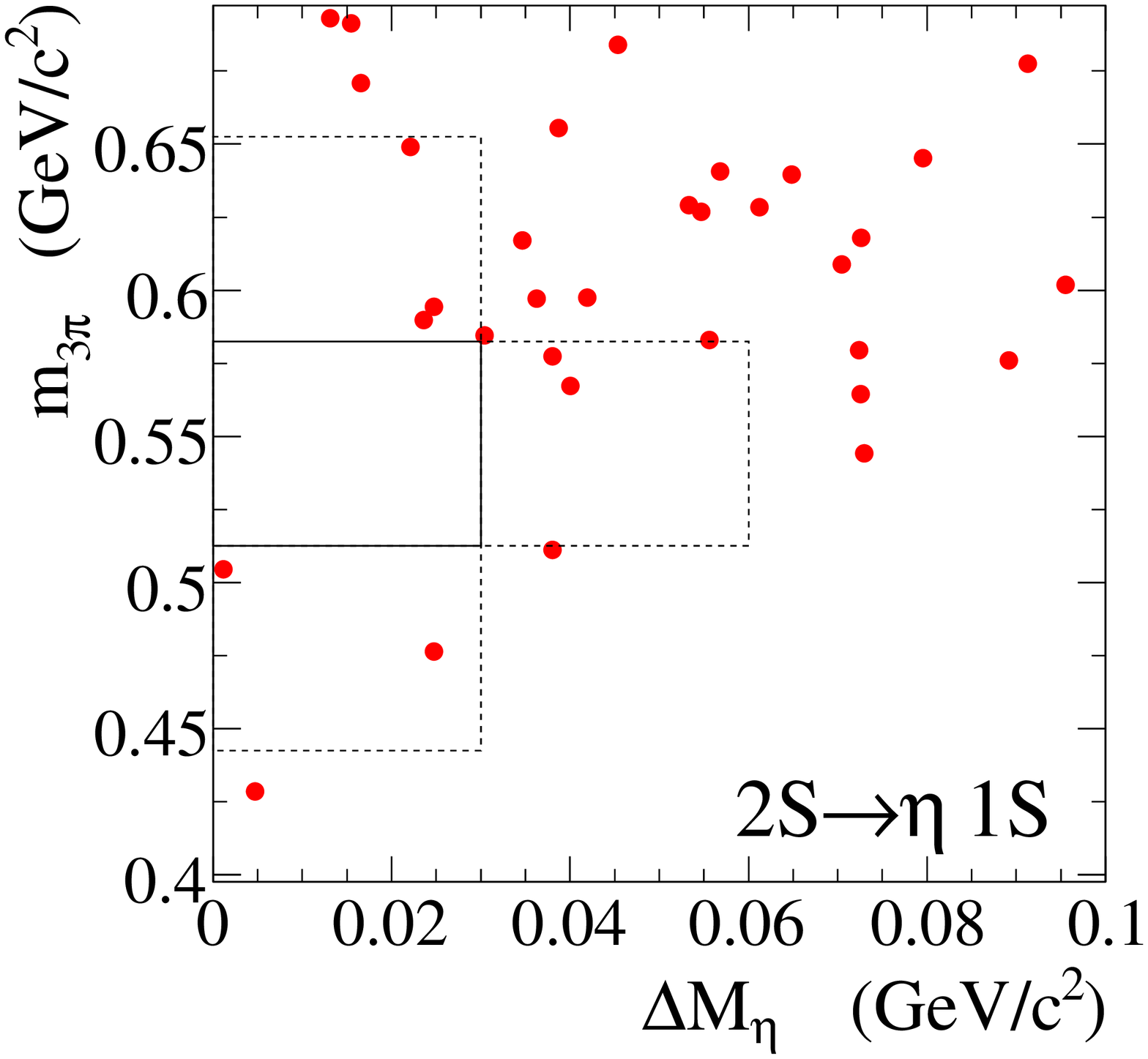}&
\includegraphics[scale=0.22]{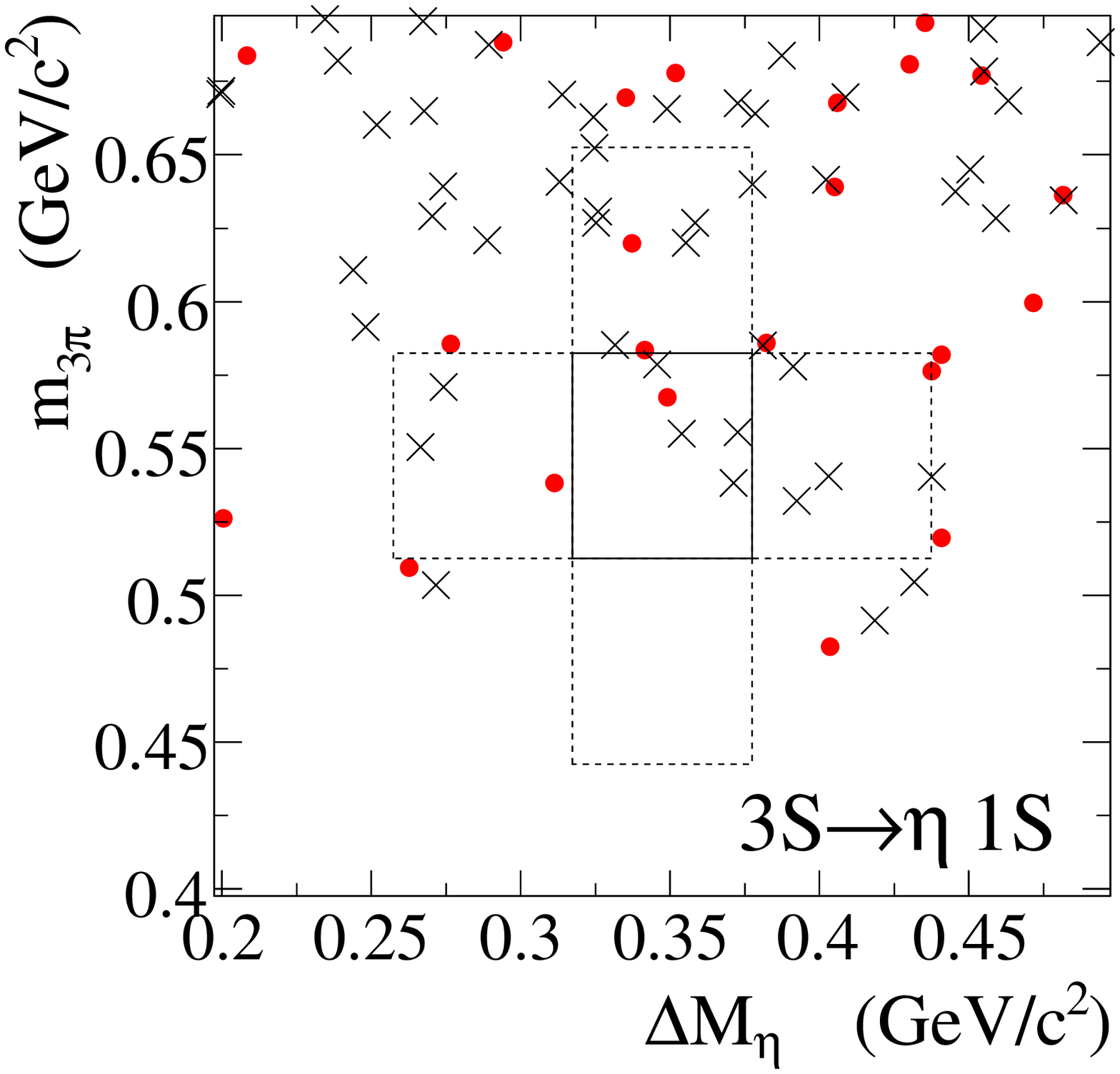}\\
\includegraphics[scale=0.22]{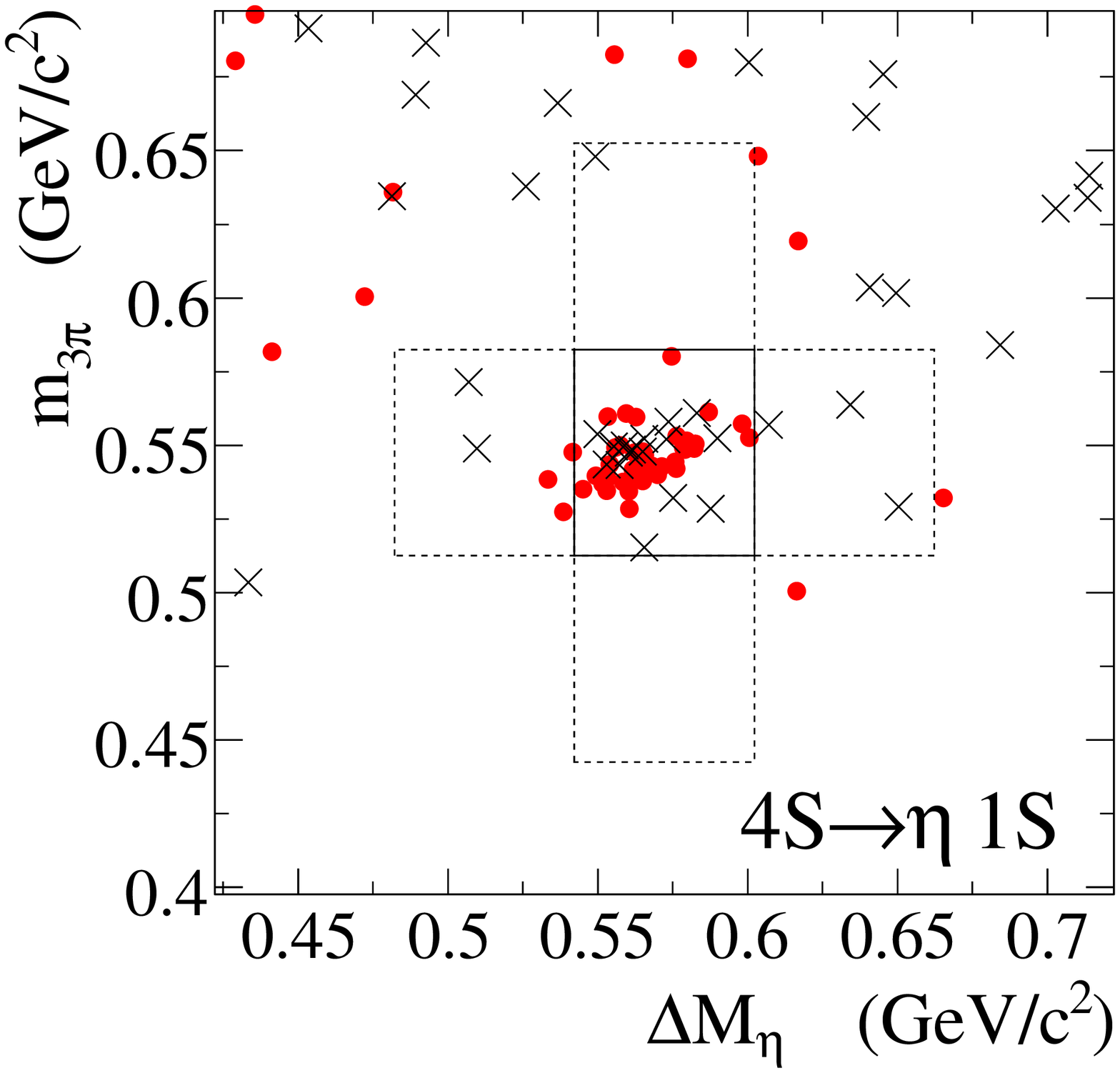}&
\includegraphics[scale=0.22]{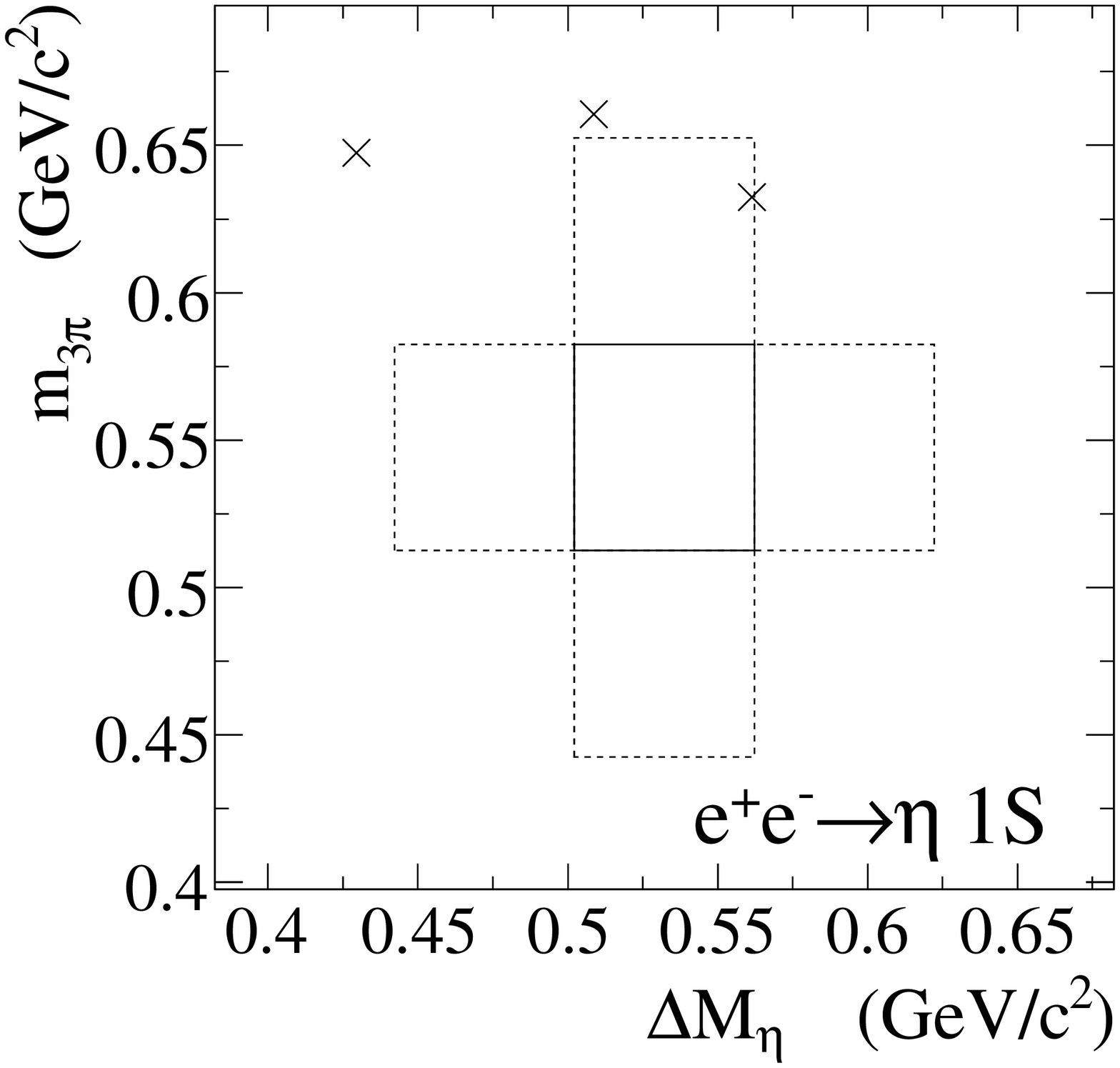}
\end{tabular}
\end{center}
\caption{\label{fig:UpsEta} Distributions of  $m_{3\pi}$ vs $\Delta M_\eta$  for the
$mS\to\eta~1S$ transitions studied.
Crosses are for the $\OneS\to\epem$ sample
and dots are for the $\OneS\to\mumu$ sample. Solid lines delimit the signal box region. 
Dashed lines delimit the sideband regions used for background extrapolation.
The signal box for the $2S\to\eta~1S$ transition (top left) is at the boundary of the kinematically
allowed region of $\Delta M_\eta$  and only one sideband  can be defined.} 
\end{figure}

\begin{figure}[!htb]
\begin{center}
\begin{tabular}{cc}
\includegraphics[scale=0.22]{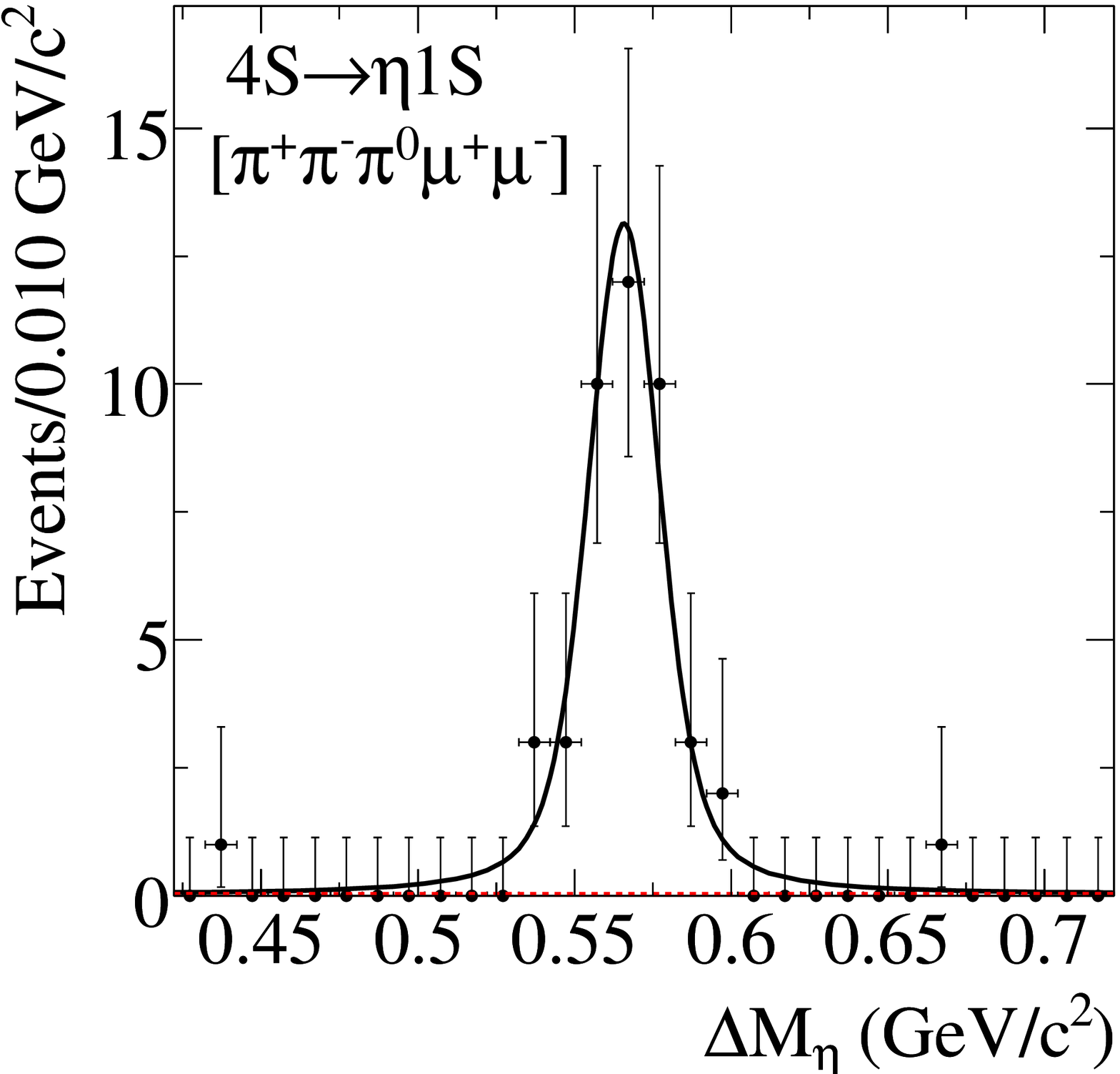}&
\includegraphics[scale=0.22]{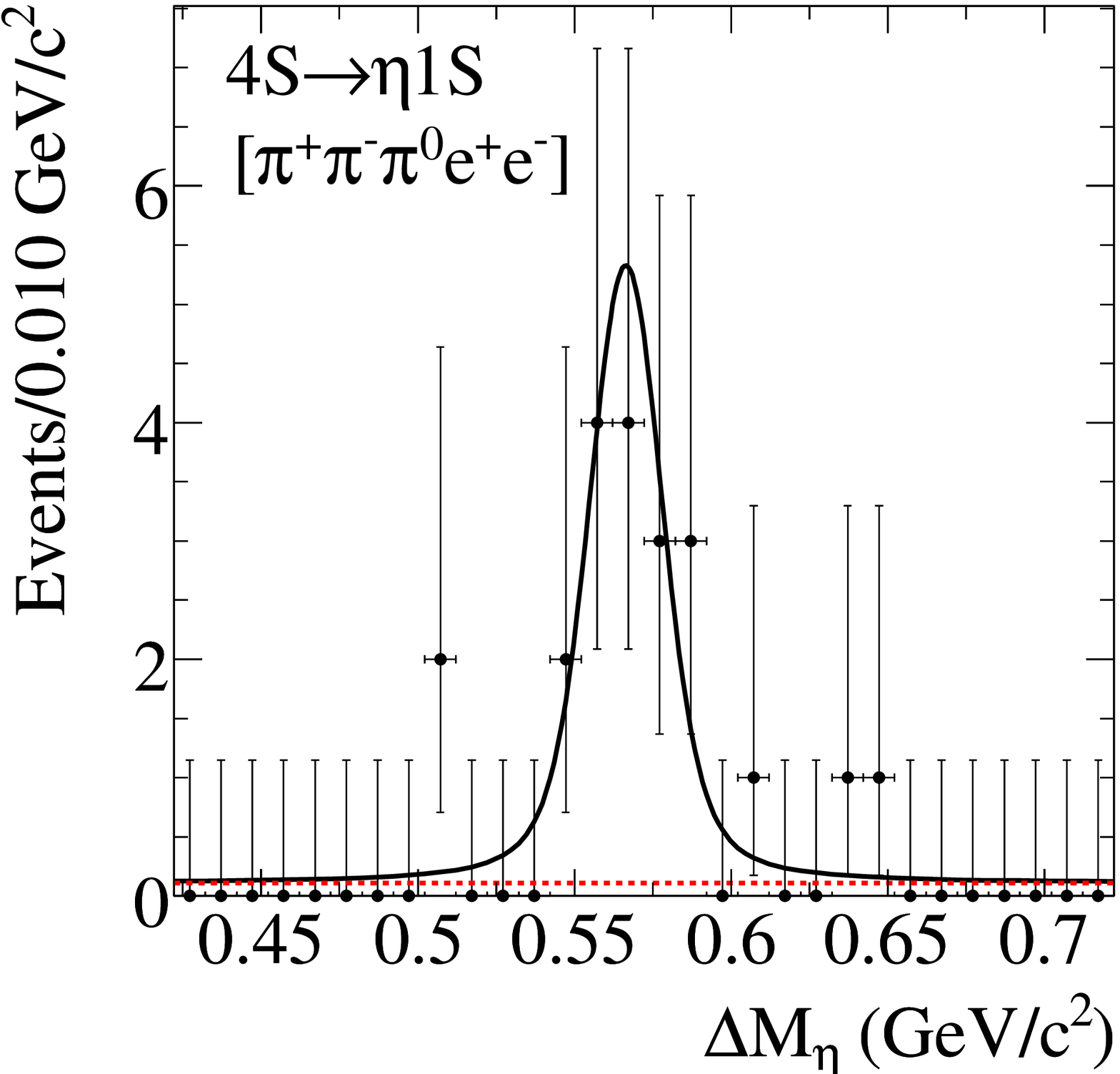}
\end{tabular}
\caption{\label{fig:UpsEtaFit}Fits to the $\Delta M_\eta$ distribution for 
$4S\to\eta~1S$ candidates with  
$\OneS\to\mumu$ (left) and  $\OneS\to\epem$ (right).
Data are shown as crosses. The solid lines show the best fit to the data. 
 Dashed lines show the background contribution.
}
\end{center}
\end{figure}

\begin{table*}[!htb]
\caption{\label{tab:FinalRes}
Results for the products of partial widths and branching fractions for the
$\Upsilon(mS)$ hadronic transitions. $N_{cand}$ is the number of
candidates in the signal box, $N_{bck}$ is the number of background events from the fit or estimated from data sidebands
as described in the text,  $N_{corr}$ is the efficiency-corrected number of signal events.
The first error is statistical,
the second is systematic. All upper limits are 90\%CL.}
\begin{tabular}{l l c | c c c }
\hline
\hline
Transition          &  & Our Measurement & $N_{cand}$  & $N_{bck}$ & $N_{corr}$ \\
\hline
$\Gamma_{ee}(2S)\times\BR{\TwoS}{\pipi\OneS}\times\BR{\OneS}{\mumu}$ &(meV) & 2582$\pm$28$\pm$94 
& 9036  & 156$\pm$11  & 24319$\pm$268  \\  
$\Gamma_{ee}(2S)\times\BR{\TwoS}{\pipi\OneS}\times\BR{\OneS}{\epem}$ &(meV) & 2618$\pm$60$\pm$97 
& 3139  & 230$\pm$9   & 25202$\pm$574  \\       
$\Gamma_{ee}(2S)\times\BR{\TwoS}{\eta\OneS}\times\BR{\OneS}{\mumu}\times\BR{\eta}{\pipi\pi^0}$ &(meV) & $<3.1$ 
& 0 & 2.5$\pm$1.1  & $<28$   \\    
$\Gamma_{ee}(3S)\times\BR{\ThreeS}{\pipi\OneS}\times\BR{\OneS}{\mumu}$ &(meV) & 457$\pm$8$\pm$18 
& 4198 & 207$\pm$10    &  9945$\pm$174  \\  
$\Gamma_{ee}(3S)\times\BR{\ThreeS}{\pipi\OneS}\times\BR{\OneS}{\epem}$ &(meV) & 441$\pm$12$\pm$18 
& 3604 & 1234$\pm$20   &  9821$\pm$261  \\  
$\Gamma_{ee}(3S)\times\BR{\ThreeS}{\pipi\TwoS}\times\BR{\TwoS}{\epem}$ &(meV) & 206$\pm$11$\pm$12
& 975  & 180$\pm$21    &  4477$\pm$241   \\  
$\Gamma_{ee}(3S)\times\BR{\ThreeS}{\eta\OneS}\times\BR{\OneS}{\mumu}\times\BR{\eta}{\pipi\pi^0}$ &(meV) & $<2.0$
& 1 & 0.8$\pm$0.4 & $<41$    \\     
$\Gamma_{ee}(3S)\times\BR{\ThreeS}{\eta\OneS}\times\BR{\OneS}{\epem}\times\BR{\eta}{\pipi\pi^0}$ &(meV) & $<9.6$
& 4 & 2.8$\pm$0.8 & $<210$    \\       
$\BR{\FourS}{\pipi\OneS}\times\BR{\OneS}{\mumu}$ &($\times 10^{-6}$) & 1.99$\pm$0.16$\pm$0.07 
& 687 &  378$\pm$11   & 739$\pm$ 60  \\  
$\BR{\FourS}{\pipi\OneS}\times\BR{\OneS}{\epem}$ &($\times 10^{-6}$) &1.76$\pm$1.05$\pm$0.06 
&1057 &  934$\pm$17   & 676$\pm$397  \\
$\BR{\FourS}{\pipi\TwoS}\times\BR{\TwoS}{\mumu}$ &($\times 10^{-6}$) & 1.65$\pm$0.21 $\pm$0.11 
& 377 & 204$\pm$8    &  615$\pm$ 78    \\   
$\BR{\FourS}{\pipi\TwoS}\times\BR{\TwoS}{\epem}$ &($\times 10^{-6}$)  &1.76$\pm$1.03 $\pm$0.11 
& 251 & 206$\pm$8    &  669$\pm$392     \\       
$\BR{\FourS}{\eta\OneS}\times\BR{\OneS}{\mumu}\times\BR{\eta}{\pipi\pi^0}$ &($\times 10^{-6}$) & 1.08$\pm$0.17$\pm$0.05 
& 40 & 0.2$\pm$0.4  & 387$\pm$60 \\       
$\BR{\FourS}{\eta\OneS}\times\BR{\OneS}{\epem}\times\BR{\eta}{\pipi\pi^0}$ &($\times 10^{-6}$)  &1.15$\pm$0.29$\pm$0.05 
& 16 & 0.7$\pm$0.6  & 424$\pm$106 \\
\hline       
\hline
\end{tabular}

\end{table*}

\subsection{\boldmath $\Upsilon(mS)\to\eta\Upsilon(1S)$}
Figure \ref{fig:UpsEta} shows the $m_{3\pi}$ vs  $\Delta M_{\eta}$ distributions for events selected
as $mS\to\eta~1S$ candidates.
The widths of the signal boxes have been chosen as $\approx \pm 3\sigma$ in 
both variables based on MC simulation: $\vert m_{3\pi}-m_\eta\vert <35\,\mevcc$ and
$\vert M(mS) -M(1S)-M(\eta) -\Delta M_{\eta}\vert <30\,\mevcc$. For the $2S\to\eta~1S$ transition
we require $\Delta M_\eta<30\,\mevcc$ because the signal for this transition is expected close to the 
kinematic limit.  

The numbers of candidates in the $2S\to\eta~1S$ and $3S\to\eta~1S$ signal boxes, shown in
Table~\ref{tab:FinalRes}, are compatible with the backgrounds extrapolated from the sidebands
defined in  Fig.~\ref{fig:UpsEta}. Thus, we have no signal for the $2S\to\eta~1S$ and $3S\to\eta~1S$ transitions. 

We observe 56 candidates for the $4S\to\eta~1S$ transition in the ``on-peak''
data sample, and no candidates in the ``off-peak'' data sample. 
We test the hypothesis that the $\epem\to\eta~1S$ cross section is the same
in the ``on-peak'' and ``off-peak'' samples by calculating the
binomial probability, ${\cal P}$, of observing respectively 56 and 0 events for a binomial coefficient of $p=L^{int}_{on}/(L^{int}_{on}+ L^{int}_{off})=0.905$,
based on the integrated luminosities of the two samples.
We obtain ${\cal P}=4\times10^{-3}$ and thus we attribute the observed $\eta\OneS$ events to \FourS\ decays.

The event yields for the $4S\to\eta~1S$ transition in the $ee$ and $\mu\mu$ final states are determined by
unbinned extended maximum likelihood fits to the $\Delta M_\eta$ distribution of the sample of events
in Fig.~\ref{fig:UpsEta} having $m_{3\pi}$ within 35~\mevcc of the known $\eta$ mass.
The signal PDF is parametrized by a Voigtian function, with resolution parameters
fixed to the values determined from MC events, while the background is assumed to be constant.
The free parameters in the fits are:
$\Delta M_{\eta~sig}$, the peak position of the signal distribution, $N_{sig}$ and $N_{bkg}$, the number of signal and background events.
The efficiency and acceptance
are determined from MC samples. The fits are shown in Fig.~\ref{fig:UpsEtaFit}.
The significance, estimated from the likelihood ratio $n\sigma\simeq\sqrt{2\log\left[{{\cal L}(N_{sig})/{\cal L}(0)}\right]}$
between a fit that includes a signal function and a fit with only a background hypothesis, is 11~$\sigma$ and 6.2~$\sigma$, respectively 
in the $\mu\mu$ and the $ee$ samples.

The 90\% CL upper limits on the signal yields for the $3S\to\eta~1S$ and $2S\to\eta~1S$ transitions are
conservatively estimated from the numbers of events in the signal boxes, taking into account the
uncertainties in the efficiencies~\cite{Barlow:2002bk}. The background level in the \mumu\ sample is
negligible, and background subtraction in the \epem\ sample, which also has a lower efficiency, would not
affect the result.

\begin{table*}[!htb]
\caption{\label{tab:SystSumm} 
Sources of systematic uncertainties on partial widths or branching fractions 
and ratios of partial widths,
separated into errors that cancel in ratios, errors due to lepton identification
(ID) and invariant mass that are common to all transitions, but differ
for electrons and muons, and errors that are specific to individual decay modes.
All errors are relative and given in percent.
We also list the corrections applied to account for differences
between data and simulation.}  
\begin{tabular}{l|c|cc|ccc|ccc} 
\hline
\hline
Source         & data/MC &\multicolumn{2}{|c|}{$\TwoS\to$}&\multicolumn{3}{|c|}{$\ThreeS\to$}&\multicolumn{3}{|c}{ $\FourS\to$} \\
               & corr.      &~~ $\pi\pi\OneS$ ~&~ $\eta\OneS$ ~~&~~ $\pi\pi\OneS$ ~&~ $\pi\pi\TwoS$ ~&~ $\eta\OneS$ ~~&~~ $\pi\pi\OneS$ ~&~ $\pi\pi\TwoS$ ~&~ $\eta\OneS$~~ \\  
\hline
\multicolumn{10}{c}{Common systematic errors (cancel in all ratios) (\%) }\\
\hline
Number of $\FourS$&       & \multicolumn{2}{|c|}{--}& \multicolumn{3}{|c|}{--} &\multicolumn{3}{|c }{1.1} \\
ISR luminosity   &       & \multicolumn{2}{|c|}{3.0} & \multicolumn{3}{|c|}{3.0}  &\multicolumn{3}{|c}{--}  \\
Tracking       &       & \multicolumn{2}{|c|}{1.0}&\multicolumn{3}{|c|}{1.0}&\multicolumn{3}{|c}{1.0}    \\
Selection      &       & \multicolumn{2}{|c|}{0.3} & \multicolumn{3}{|c|}{0.3} & \multicolumn{3}{|c}{0.3}            \\
$p^*_{cand}$ cut &     & \multicolumn{2}{|c|}{0.3} & \multicolumn{3}{|c|}{0.3} & \multicolumn{3}{|c}{0.3}            \\
\hline
\multicolumn{10}{c}{Systematic errors associated to lepton identification or invariant mass (\%)}\\
\hline
Muon ID        & 1.025 & \multicolumn{2}{|c|}{0.6} & \multicolumn{3}{|c|}{0.6} & \multicolumn{3}{|c}{0.6}  \\
$M(\mumu)$ cut   & 1.006 & \multicolumn{2}{|c|}{0.2} & \multicolumn{3}{|c|}{0.2} & \multicolumn{3}{|c}{0.2}  \\
Electron ID         & 1.011 & \multicolumn{2}{|c|}{0.7} & \multicolumn{3}{|c|}{0.7} & \multicolumn{3}{|c}{0.7}  \\ 
$M(\epem)$ cut   & 0.998 & \multicolumn{2}{|c|}{0.5} & \multicolumn{3}{|c|}{0.5} & \multicolumn{3}{|c}{0.5}  \\
\hline
\multicolumn{10}{c}{Systematic errors specific to each mode  (\%)}\\
\hline
$\pi^0$ efficiency   & 1.033 & --  & 3.6                & --  & --  & 3.6           & --  & --  & 3.6 \\
Acceptance     &       & 0.3& --                 & 1.7 & 4.7 & --            & 2.6 &  6.0 & -- \\
Fitting        &       & 1.6 & 1.6                & 1.6 & 1.6 & 1.6           & 1.6 & 1.6 & 1.6 \\  
\hline
Total \epem (\%)   &       & 3.7 & --                 & 4.1 & --  & 5.1           & 3.5 & 6.5 & 4.4 \\
Total \mumu (\%)   &       & 3.7 & 5.1                & 4.0 & 5.9 & 5.1           & 3.5 & 6.5 & 4.3 \\
\hline
Total on ratios (\%) &      & --  & 4.3                & --  & 5.5 & 4.6           & --  & 6.9 & 5.0 \\
\hline
\hline  
\end{tabular}

\end{table*}

\section{\boldmath Systematic Uncertainties}
\label{sec:sys}

We have considered a number of possible sources of systematic uncertainties, in addition to the
number of \FourS~\cite{Aubert:2002hc} and the calculated luminosity for ISR events.
The uncertainties in charged  track  and $\pi^0$ reconstruction efficiencies are determined by a comparison 
of data and MC events on independent control samples.  
The systematic uncertainties associated with the event selection, 
the cut on  $p^*_{cand}$, the $M_{\ell\ell}$ invariant mass cut,
and the lepton identification criteria  are estimated by comparing 
the efficiencies determined from MC samples to the corresponding efficiencies
measured with the ISR $mS\to\pi\pi\;nS$ samples in the modes where there are sufficiently high statistics and low background to allow the comparison.
The efficiencies are determined from the numbers of signal events which pass or 
fail any given cut, after all other cuts are applied. 

The systematic uncertainties due to the choice of signal and background parameterizations  are estimated
by using different functions or different parameters, and by varying the $\Delta M$ or $\Delta M_\eta$ fit ranges.
The uncertainty in the acceptance correction for the $mS\to\pi\pi\; nS$ transitions
is determined by the change in the signal yields when using different $m_{\pi\pi}$ and $\cos{\theta_{h}}$ binnings.

The systematic uncertainties from all these sources 
are summarized in Table~\ref{tab:SystSumm} for each transition. 
The total systematic uncertainty is
estimated by adding in quadrature all different contributions.
We apply correction factors to the efficiency determined from MC events, accounting 
for differences between data and MC samples in
the $\pi^0$ reconstruction,  in lepton identification, and
in the $M_{\ell\ell}$ cut.   

\begin{table*}[!tb]
\caption{\label{tab:Ratios}Our measurements for the products and ratios of partial widths and branching fractions   
of $\Upsilon(mS)$ hadronic transitions, with comparisons to previous measurements and theoretical expectations.
We also report the values of the branching fractions that are derived from our measurements using world
average values for  $\Gamma_{ee}(nS)$.  All upper limits are 90\% CL.
The values of the last seven branching fractions in this Table (reported below the horizontal line) are not independent from the values reported above.
The values of ${\cal B}(\FourS\to\pipi\Upsilon(nS))$ from Ref.~\cite{PDG07} and indicated with an asterisk 
are based on our previous measurement~\cite{Aubert:2006bm} performed on a subset of the current
sample. As discussed in the text, part of the difference in the central values
is ascribed to a more accurate estimate of the acceptance.}
\begin{tabular}{l l|c c c }
\hline
\hline
\hbox to 6.0truecm{\null\hfill}   &  ~~~~~~~    &~~~~~~~ This work  ~~~~~~~&~~ PDG~\cite{PDG07} ~~&~~ Prediction~~  \\
\hline

$\Gamma_{ee}(2S)\times\BR{\TwoS}{\pipi\OneS}$ & (\ev)  & 105.4$\pm$1.0$\pm$4.2 & 115$\pm$5 & \\
${\GG{\TwoS}{\eta\OneS}}/{\GG{\TwoS}{\pipi\OneS}}$ & ($\times10^{-3}$) 
& $<5.2$ & $<11$ &2.5~\cite{Kuang-Rev}  \\
$\Gamma_{ee}(3S)\times\BR{\ThreeS}{\pipi\OneS}$ & (\ev) & 18.46$\pm$0.27$\pm$0.77 &19.8$\pm$1.0 & \\
${\GG{\ThreeS}{\pipi\TwoS}}/{\GG{\ThreeS}{\pipi\OneS}}$  ~~~&       
&~ 0.577$\pm$0.026$\pm$0.060 ~& 0.63$\pm$0.14 &0.3~\cite{Kuang-Rev}  \\
${\GG{\ThreeS}{\eta\OneS}}/{\GG{\ThreeS}{\pipi\OneS}}$ ~~~&($\times10^{-2}$)
& $<1.9$ & $<5$ &1.7~\cite{Kuang-Rev}  \\
${\cal B}({\FourS}\to{\pipi\OneS})$ &($\times10^{-4}$) ~~
& 0.800$\pm$0.064$\pm$0.027 & 0.90$\pm$0.15$^{(\ast)}$ & -- \\
${\GG{\FourS}{\pipi\TwoS}}/{\GG{\FourS}{\pipi\OneS}}$ &
&1.16$\pm$0.16$\pm$0.14  &  & -- \\
${\GG{\FourS}{\eta\OneS}}/{\GG{\FourS}{\pipi\OneS}}$   & 
& 2.41$\pm$0.40$\pm$0.12 & -- & -- \\
\hline
${\cal B}(\TwoS\to\pipi\OneS)$ &(\%) & 17.22$\pm$0.17$\pm$0.75  & 18.8$\pm$0.6 & 27$\pm$2~\cite{Kuang-Rev} \\
${\cal B}(\TwoS\to\eta\OneS)$ &($\times 10^{-4}$) & $<9$  & $<20$ & 8.1$\pm$0.8~\cite{Eichten:2007qx} \\
${\cal B}(\ThreeS\to\pipi\OneS)$ & (\%) & 4.17$\pm$0.06$\pm$0.19  & 4.48$\pm$0.21 & 3.3$\pm$0.3~\cite{Kuang-Rev}\\
${\cal B}(\ThreeS\to\pipi\TwoS)$ & (\%) & 2.40$\pm$0.10$\pm$0.26  & 2.8$\pm$0.6 & 1.0$\pm$0.1~\cite{Kuang-Rev}\\
${\cal B}(\ThreeS\to\eta\OneS)$ & ($\times 10^{-4}$) & $<8$  & $<22$ & 6.7$\pm$0.7~\cite{Eichten:2007qx} \\
${\cal B}(\FourS\to\pipi\TwoS)$ &($\times 10^{-4}$) & 0.86$\pm$0.11$\pm$0.07  & 0.88$\pm$0.19$^{(\ast)}$ & -\\ 
${\cal B}(\FourS\to\eta\OneS)$ &($\times 10^{-4}$) & 1.96$\pm$0.06$\pm$0.09 & -- &--\\
\hline
\hline
\end{tabular}
\end{table*}

\section{\boldmath Results}
\label{sec:res}
The products of branching fractions and partial widths for each transition 
are given Table~\ref{tab:FinalRes}. They are determined from the efficiency-corrected yield in each mode, 
after correcting for small differences between data and MC samples and taking into account 
the number of $\FourS$ or the equivalent ISR luminosity, ${\cal K}$.
For a narrow vector resonance produced in ISR
\begin{equation} {\cal K}={ L^{int}_{on}}  \frac{12\pi^2}{M(mS) \, s}
  W\left(s,1-\frac{M^2(mS)}{s}\right) 
\end{equation} 
where the QED ``radiator'' function  $W(s,x)$ is calculated to second order 
following~\cite{Benayoun:1999hm,Bonneau:1971mk,Baier:1973ms}.

Averaging the results from the $\epem$ and the $\mumu$ final states, taking into account the common systematic errors, 
and using the world average values of
${\cal B}(\eta\to\pi^+\pi^-\pi^0)$ and ${\cal B}(\Upsilon(nS)\to\ellell)$~\cite{PDG07} 
we obtain the partial widths and ratios of partial widths listed in Table~\ref{tab:Ratios}.
In this Table, we also compare our results to the values expected for each quantity based on
previous measurements of $\Upsilon(mS)$ widths and branching fractions. The measured values of the
\TwoS\ and \ThreeS\ total widths
are  used to derive the theoretical expectations for branching fractions
from the predicted partial widths in~\cite{Kuang-Rev}.

The values of ${\cal B}(\FourS\to\pipi\OneS)\times{\cal B}(\OneS\to\mumu)$ and
${\cal B}(\FourS\to\pipi\TwoS)\times{\cal B}(\TwoS\to\mumu)$
supersede our previously reported values based on a fraction of the current sample~\cite{Aubert:2006bm}.
Part of the difference in the central values is due to the different methods used to determine 
the acceptance, which was calculated in our previous paper assuming  
a phase-space distribution in the $\FourS\to\pipi\Upsilon(nS)$
decay.  The efficiency is not uniform over the Dalitz plot, thus the
impact on the central value between the two methods depends on the angular
distributions peculiar of each transition. The difference can be
estimated by comparing the value of the phase-space averaged efficiencies
$\varepsilon_{PS}$,  and the effective efficiencies
$\varepsilon_{eff}$ calculated from the observed event yields in each 
region of the Dalitz plot
\begin{equation}
\varepsilon_{eff}=\frac{\sum_{i=1}^{nbins}N_{sig}^i}{\sum_{i=1}^{nbins}N_{sig}^i/\varepsilon_i}.\label{effdef}
\end{equation} 
Notice that the uncertainty in the calculated effective efficiency is due to the
statistical uncertainty in the event yield.
As shown in Table~\ref{tab:effs} the effective efficiency for 
$\FourS\to\pipi\OneS$, when the \OneS\ decays to \mumu, is $\sim 7\%$ larger than
the value estimated using a phase-space distribution. 
Accounting for this difference, the results presented here are statistically compatible with
the ones previously reported.

From our result we derive new values for  ${\cal B}(\Upsilon(3S,2S)\to\pipi\Upsilon(1S))$ that 
are of comparable precision to the previous world averages, and compatible with them.
The value of  ${\cal B}(\ThreeS\to\pipi\TwoS)$ derived from our measurement has an error
that is smaller than the current world average.

\section{\boldmath Conclusions}
\label{sec:concl}
We have presented a study of hadronic transitions between the $\Upsilon$ states:
new measurements of the branching fractions ${\cal B}(\FourS\to\pipi\Upsilon(1S,2S))$, ${\cal B}(\ThreeS\to\pipi\Upsilon(2S))$
which have smaller errors than current world averages, 
and new measurements of ${\cal B}(\Upsilon(3S,2S)\to\pipi\OneS)$ whose precision is comparable to
present world averages. We have also presented measurements of the ratios of partial widths
$\GG{\Upsilon(mS)}{\pipi\Upsilon(2S)}/\GG{
\Upsilon(mS)}{\pipi\Upsilon(1S)}$ ($m=3,4$) where a number of systematic uncertainties cancel.
Our results for the branching fractions of the $\TwoS$ and $\ThreeS\to\eta\OneS$ transitions 
represent improvements over the current published upper limits, and are compatible with the recent results from CLEO~\cite{He:2008xk}:  ${\cal B}(\TwoS\to\eta\OneS)=
(2.1^{+0.7}_{-0.6}\pm0.5)\times10^{-4}$,  ${\cal B}(\ThreeS\to\eta\OneS)<2.9\times10^{-4}$ at 90\% CL.

We observe a significant number of $\eta\OneS$ candidates at the formation energy of the $\FourS$.
We can exclude the hypothesis that they are due to continuum $\epem\to\eta\OneS$ with a probability of 99.6\%
and we attribute them to \FourS\  decays.
The branching fraction for the $\FourS\to\eta\OneS$ decay is  larger
than the branching fraction for $\FourS\to\pipi\OneS$, which is unexpected when compared to all other known charmonium
and bottomonium transitions.  
There are no predictions for this specific decay mode.
In the QCDME calculation for hadronic transitions, the effect of the nodes in the wave functions
in the overlap integrals between the initial and final states and the intermediate states can be large
for radial excitations. 
But even that should not significantly affect the ratio of partial widths 
$\Gamma(\FourS\to\eta\OneS)/\Gamma(\FourS\to\pipi\OneS)$,
at least if the $\FourS\to\eta\OneS$ transition is E1M2~\cite{Kuang-Rev}.
It is possible that accidental cancellations suppress the E1M2 term with respect 
to M1M1, or perhaps QCDME becomes unreliable for higher gluon momenta.
These results, together with the recent CLEO measurement of 
the matrix elements in $\ThreeS\to\pipi\Upsilon(1S,2S)$ and $\TwoS\to\pipi\OneS$
transitions~\cite{CroninHennessy:2007zz}, 
could provide a tool to understand the hadronic transitions better.

\section{Acknowledgments}
We are grateful for the 
extraordinary contributions of our \pep2\ colleagues in
achieving the excellent luminosity and machine conditions
that have made this work possible.
The success of this project also relies critically on the 
expertise and dedication of the computing organizations that 
support \babar.
The collaborating institutions wish to thank 
SLAC for its support and the kind hospitality extended to them. 
This work is supported by the
US Department of Energy
and National Science Foundation, the
Natural Sciences and Engineering Research Council (Canada),
the Commissariat \`a l'Energie Atomique and
Institut National de Physique Nucl\'eaire et de Physique des Particules
(France), the
Bundesministerium f\"ur Bildung und Forschung and
Deutsche Forschungsgemeinschaft
(Germany), the
Istituto Nazionale di Fisica Nucleare (Italy),
the Foundation for Fundamental Research on Matter (The Netherlands),
the Research Council of Norway, the
Ministry of Education and Science of the Russian Federation, 
Ministerio de Educaci\'on y Ciencia (Spain), and the
Science and Technology Facilities Council (United Kingdom).
Individuals have received support from 
the Marie-Curie IEF program (European Union) and
the A. P. Sloan Foundation.

\end{document}